\documentclass[11pt]{article}
\usepackage[a4paper]{geometry}
\usepackage[T1]{fontenc}
\usepackage[latin9]{inputenc}
\usepackage{float}
\usepackage{amssymb}
\usepackage{amsmath}
\usepackage{graphicx}
\usepackage{esint}
\usepackage{mathtools}% http://ctan.org/pkg/mathtools
\makeatletter
\newcommand{\mytag}[2]{%
  \text{#1}%
  \@bsphack
  \protected@write\@auxout{}%
         {\string\newlabel{#2}{{#1}{\thepage}{\relax}{\relax}}}%
  \@esphack
}

\usepackage{color}
\definecolor{gray}{gray}{0.5}

\usepackage[caption=false]{subfig}
\makeatletter
\@ifundefined{textcolor}{}
{%
 \definecolor{BLACK}{gray}{0}
 \definecolor{WHITE}{gray}{1}
 \definecolor{RED}{rgb}{1,0,0}
 \definecolor{GREEN}{rgb}{0,1,0}
 \definecolor{BLUE}{rgb}{0,0,1}
 \definecolor{CYAN}{cmyk}{1,0,0,0}
 \definecolor{MAGENTA}{cmyk}{0,1,0,0}
 \definecolor{YELLOW}{cmyk}{0,0,1,0}
 }

\usepackage{graphicx}  % needed for figures
\usepackage{dcolumn}   % needed for some tables
\usepackage{bm}        % for math
\usepackage{amssymb}   % for math

\usepackage{color}

\usepackage{afterpage}
\newcommand{\bl}{\relax}
\newcommand{\el}{\relax}

\newcommand{\eps}{\varepsilon}

\newcommand{\ti}[1]{\tilde{#1}}
\renewcommand{\ti}[1]{{#1}}

\newcommand{\chefab}{(\ref{chefb}a,b)}

\makeatother

\usepackage[english]{babel}

\begin{document}

\title{Sharp Interface Limits of the Cahn--Hilliard Equation with Degenerate Mobility}

\author{Alpha Albert Lee, Andreas M\"unch, Endre S\"uli
\thanks{Mathematical Institute,
University of Oxford,
Andrew Wiles Building,
Woodstock Road,
Oxford,
OX2 6GG}}

\maketitle

\begin{abstract}
In this work, the sharp interface limit of the degenerate
Cahn--Hilliard equation (in two space dimensions) with
a polynomial double well free energy and a quadratic
mobility is derived via a matched asymptotic analysis
involving exponentially large and small terms and multiple
inner layers. In contrast to some results found in the
literature, our analysis reveals that the interface motion
is driven by a combination of surface diffusion flux
proportional to the surface Laplacian of the interface
curvature and an additional contribution from nonlinear,
porous-medium type bulk diffusion, For higher degenerate
mobilities, bulk diffusion is subdominant. The sharp
interface models are corroborated by comparing relaxation
rates of perturbations to a radially symmetric stationary
state with those obtained by the phase field model.
\end{abstract}

%%%%%%%%%%%%%%%%%%%%%%%%%%%%%%%%%%%%%%%%%%%%%%%%%%%%%%%%%%%%%%%%%%%%%%%

\section{Introduction} 

Phase field models are a common framework to describe the
mesoscale kinetics of phase separation and pattern-forming processes
\cite{provatas2010phase, chen2002phase}. Since phase field models
replace a sharp interface by a diffuse order parameter profile, they
avoid numerical interface tracking, and are versatile enough to capture
topological changes.  Although such models can be constructed starting
from a systematic coarse-graining of the microscopic Hamiltonian
\cite{giacomin2000macroscopic,giacomin1998phase,giacomin1997phase,
giacomin1996exact}, the use as a numerical tool to approximate 
a specific free boundary problem requires in the first instance careful
consideration of their asymptotic long-time sharp interface limits.

In this paper, we will mainly focus on the Cahn-Hilliard equation  
for a single conserved order parameter $u=u(\mathbf{x},t)$,
\begin{subequations}\label{main}
\begin{equation}
u_t = - \nabla \cdot \mathbf{j}, 
\qquad 
\mathbf{j} = - M(u) \nabla \mu
\qquad
\mu = - \eps^2 \nabla^2 u + f'(u).  
\label{CHE}
\end{equation}
with
a double well potential 
\bl
\begin{align}
f(u) &=(1-u^2)^2/2
\label{mainf}
\intertext{%
and the degenerate, quadratic mobility
}
M(u)&=(1-u^2)_+, 
\label{mainm}
\intertext{on a bounded two-dimensional domain $\Omega$ with boundary conditions}
\nabla u\cdot\mathbf{n}
 = 0, &\qquad \mathbf{j}\cdot\mathbf{n} = 0
\label{BC_usual}
\end{align}
\el
\end{subequations}
at $\partial \Omega$. Here, $(\cdot)_+$ is the positive part of the
quantity in the brackets, $\mathbf{x}$ represents the two-dimensional
spatial coordinates, $t$ is the time, $\mu$ the chemical potential,
$\mathbf{j}$ the flux, and $\mathbf{n}$ the outward pointing normal
to $\partial \Omega$. Boldface characters generally represent
two-dimensional vectors.  Both the potential and the mobility are
defined for all $u$.
The mobility is continuous but not differentiable 
at $u=\pm 1$.

The case of a Cahn-Hilliard equation with a constant mobility 
has been intensively discussed in the literature. 
In particular, the sharp interface limit $\eps\to0$ was determined
by Pego \cite{pego1989front}, and subsequently proven rigorously
by Alikakos et al.~\cite{alikakos1994convergence}. On a long
time scale, $t=O(\eps^{-1})$, the result is the
Mullins--Sekerka problem \cite{mullins1963morphological}. 
In particular, the motion of the interface
between the two phases is driven by flux from bulk diffusion.

In contrast, Cahn-Hilliard equations with degenerate mobility are commonly
expected to approximate interface motion by surface diffusion~\cite{Mulli57} on the time
scale $t=O(\eps^{-2})$, where the interface velocity $v_n$ is proportional to the surface Laplacian $\Delta_s$
of the interface curvature $\kappa$,
\begin{equation}\label{vnsl}
v_n\propto\Delta_s\kappa.
\end{equation}
We note that the surface Laplacian is equal to $\partial_{ss}\kappa$ in two space dimensions, where $s$ 
is the arclength.
In fact, for the case of the degenerate mobility
$M(u)=1-u^2$ and either the logarithmic free energy
\bl
\begin{align*}
f(u) &= 
\frac12\theta \left[(1+u)\ln(1+u) + (1-u)\ln(1-u)\right]
+\frac 12 (1-u^2),
\end{align*}
\el
with temperature $\theta=O(\eps^\alpha)$, or the
double obstacle potential
\[
f(u)=1-u^2 \quad \text{for } |u|\leq 1, 
\quad f(u)=\infty \quad \text{otherwise}, 
\]
Cahn \emph{et al.}~\cite{cahn1996cahn} showed via asymptotic expansions that the sharp interface limit is indeed 
interface motion by surface diffusion \eqref{vnsl}.

Although the logarithmic potential and the double obstacle
potential as its deep quench limit are well motivated, in particular for
binary alloys, \cite{CahnH71,CahnT94,TayloC94,CahnH58,giacomin1996exact,
kltahara1978kinetic, PuriBL97, bhate2000diffuse}, other combinations of
potentials and mobility have been used in the literature as a basis for
numerical approaches to surface diffusion~\cite{CenicG13}. Those models are often
employed in more complex situations with additional physical effects, such as the
electromigration in metals \cite{mahadevan1999phase}, heteroepitaxial
growth \cite{ratz2006surface}, anisotropic fields \cite{TorabL12,TorabLVW09}, phase separation
of polymer mixtures \cite{WolteBP06, GemmeBP05} and
more recently in solid-solid dewetting \cite{jiang2012phase} and coupled fluid
flows \cite{AbelsR09, SibleNK13,AbelsGG12}. In those models, a smooth polynomial double-well
free energy is used in combination with the mobility $M(u)=1-u^2$ or the degenerate biquadratic 
mobility
$M(u)=(1-u^2)^2$ for $|u|\leq 1$. A smooth free energy is numerically more convenient to implement, especially in a multiphyscial model, as it avoids 
the singularity present in either the logarithmic or double obstacle potential. Authors typically justify their
choice of mobility and free energy by adapting the asymptotic analysis by Pego~\cite{pego1989front}
and Cahn et al.~\cite{cahn1996cahn} to obtain the interface motion
\eqref{vnsl} in the sharp interface limit.

Interestingly, Gugenberger et al.~\cite{gugenberger2008comparison},
recently revisited some of these models and pointed out an apparent inconsistency that
appears in the asymptotic derivations except when the interface is flat.
Other evidence suggests that the inconsistency may not be a mere technicality
but that some bulk diffusion is present and enters the interfacial mass
flux at the same order as surface diffusion.  This was observed for
example by Bray and Emmott \cite{BrayE95} when considering the coarsening
rates for dilute mixtures, and by Dai and Du \cite{DaiD12} where the
mobility is degenerate on one but is constant on the other side of the interface; the papers by Glasner \cite{Glasn03} and Lu et al.~\cite{LuGBK07} also
use a one-sided degenerate mobility but consider a time regime where
all contributions from the side with the degeneracy are dominated by
bulk diffusion from the other.) In fact,
an early publication by Cahn and Taylor \cite{CahnT94}
remarked that using a biquadratic potential might not drive the order
parameter close enough towards $\pm 1$ to sufficiently suppress
bulk diffusion, citing unpublished numerical results.
Diffuse interface models for binary fluids with  a double well 
potential and a quadratic
mobility $M(u)=1-u^2$ or $M(u)=(1-u^2)_+$ are investigated in
\cite{AbelsGG12,SibleNK13}. However, in both studies, the leading order
expressions for the interface motion do not contain bulk
diffusion contributions.

In this paper, we aim to resolve the apparent conundrum in the literature, and revisit the sharp interface limit for
\eqref{main}. We will obtain a sharp interface model where the interface
motion is driven by surface diffusion, \emph{i.e.}\ the surface Laplacian,
\emph{and} a flux contribution due to nonlinear bulk diffusion either
from one or both sides of the interface, depending on 
the nature of the solutions for $u$ in the outer regime.
The matched asymptotic analysis is rather subtle, and
involves the matching of exponentially
large and small terms and multiple inner layers.

The paper is organised as follows: Section~2 approximates
solutions of \eqref{main} which satisfy $|u|\leq 1$; 
Section~3 considers the asymptotic structure of the radially symmetric
stationary state, which demonstrates the matched asymptotic expansion
and exponential matching technique in a simpler setting; Section~4
returns to the general 2D time dependent problem;
Section~5 briefly discusses the sharp interface limit for a class of
solutions with the mobility $M(u)=|1-u^2|$ where $|u|\leq 1$ is not
satisfied, and for the Cahn-Hilliard model with a biquadratic degenerate
mobility $M(u)=((1-u^2)_+)^2$; Section~6 summarises and concludes the work.

%%%%%%%%%%%%%%%%%%%%%%%%%%%%%%%%%%%%%%%%%%%%%%%%%%%%%%%%%%%%%%%%%%%%%%%%%%%%%%%%%%%%%%%%%%%%%%%

\section{Preliminaries}\label{sec:prelim}

In this paper, we are interested in the behaviour of solutions to \eqref{CHE}
describing a system that has separated into regions where $u$ is close to $\pm 1$,
except for inner layers of width $\eps$ between them, and
evolve on the typical time for surface diffusion, $t=O(\eps^{-2})$.
We thus rescale time via $\tau = \eps^2 t $, so that the
Cahn--Hilliard equation reads
\begin{subequations}\label{chefb}
\bl
\begin{equation}
\eps^2 \partial_\tau u = \nabla \cdot \mathbf{j},
\qquad \mathbf{j}= M(u) \nabla \mu,
\qquad
\mu = - \eps^2 \nabla^2 u + f'(u), \label{chereb}  
\end{equation}
\el
and we keep the boundary conditions on $\partial \Omega$,
\bl
\begin{equation}
\nabla u\cdot\mathbf{n}
 = 0, \qquad \mathbf{j}\cdot\mathbf{n} = 0.
\qquad \text{at } \partial\Omega.%\label{chefbd}
\end{equation}
\el  
We will denote the subsets where $u>0$ and $u<0$ by $\Omega_+$ and
$\Omega_-$, respectively, and identify the location of the interface
with $u=0$. Moreover, we assume that $\Omega_+$
is convex unless otherwise stated, and has $O(1)$ curvature everywhere. 
We will focus on solutions of \chefab{} that satisfy $|u|\leq 1$.
The existence of such solutions has been shown by Elliott and Garcke \cite{EllioG96}.

The general procedure to obtain a description of the interface evolution is then to consider and match expansions of \chefab,
the so-called outer expansions, with inner expansions using
appropriate scaled coordinates local to the interface. The approach
assumes that the solution of \chefab{} is quasi-stationary \emph{i.e.}\
close to an equilibrium state.  Unfortunately, it is not obvious what
the appropriate nearby equilibrium state could be in the situation
we consider here. The problem arises because equilibrium solution
to \chefab{} with constant $\mu$ does not generally satisfy the bound $|u|<1$
inside of $\Omega_+$~\cite{pego1989front}.

It is helpful to revisit the standard matched
asymptotics procedure for \chefab{} to understand the implications of
this observation. Notice that the time derivatives drop out of the
lower order outer and inner problems. The leading order inner solution
for the double well potential is simply a tanh-profile, which matches
with $\pm 1$ in the outer solution; the corresponding leading order chemical potential is zero. To
next order, the inner chemical potential is proportional to $\kappa$,
and this supplies boundary conditions for the chemical potential in
the outer problem via matching to be $\mu_1=c_1\kappa$. Here, $\mu_1$
denotes the first non-trivial contribution to the chemical potential in
the outer expansion, $\mu=\eps\mu_1+O(\eps^2)$, and $c_1$ represents
a fixed numerical value.  It is obtained from a detailed calculation
along the lines of section~\ref{sec:axi},
which in fact shows that $c_1>0$.
It is easy to see from the third equation in \eqref{chereb} that the outer correction $u_1$
for $u=\pm 1 + \eps u_1$ is given by $u_1=\mu_1/f''(\pm 1)$, thus
$u=\pm 1 + c_1 \kappa \eps/4 + O(\eps^2)$ near the interface. Inside
$\Omega_+$, we therefore have that the outer solution $u>1$.  Notice
that we have used that $f$ is smooth at $u=\pm 1$ ---  for
the double obstacle potential, there is no correction to $u=\pm 1$
in the outer problem, see \cite{cahn1996cahn}.

%%%%%%% First Figure (a) + (b) %%%%%%%%%%%% 

\begin{figure}[tp]
\centering
\includegraphics[width=0.78\textwidth]{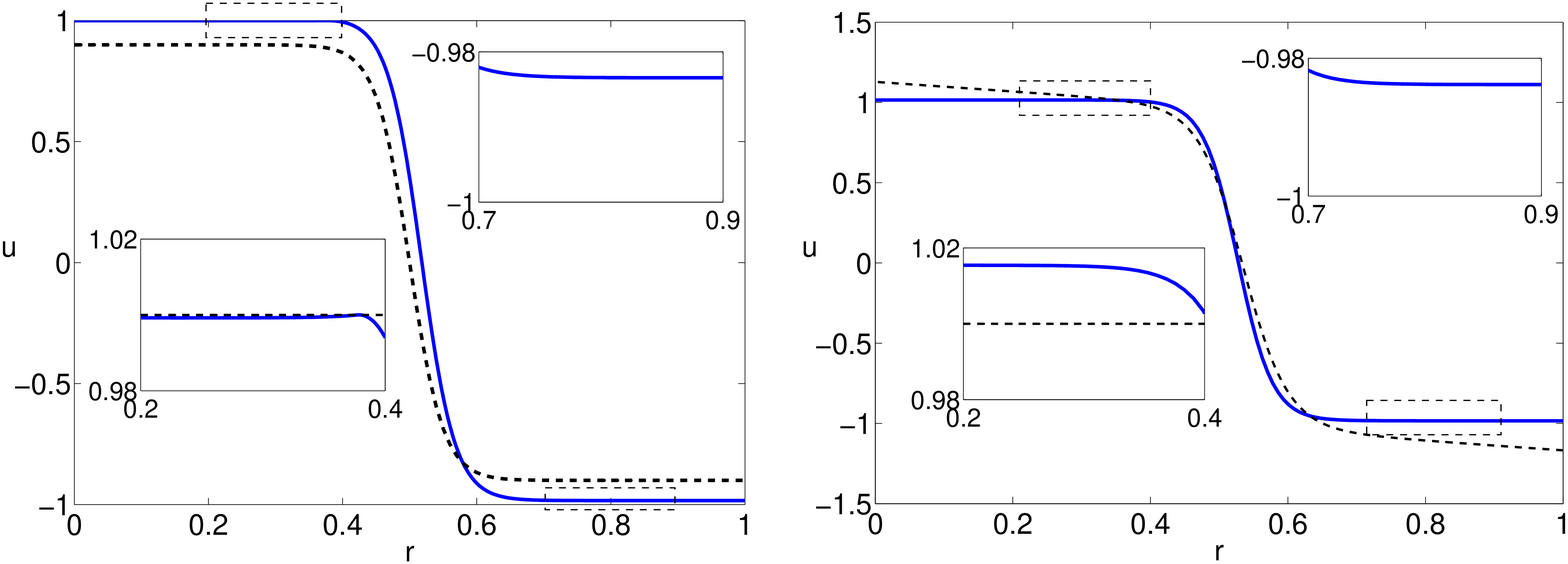}
\caption{The long-time solution $u$ for the 
radially symmetric degenerate Cahn--Hilliard
equation  \eqref{main} for different
initial data and different mobilities. In (a, left panel), the 
mobility is \eqref{mainm} and initial data is bounded within
$[-1,1]$, while in (b, right panel) it exceeds 1 and $-1$ to the left
and right, and the mobility is replaced by $M(u)=|1-u^2|$, respectively. 
In both panels, the initial data is shown by dashed lines
while the long-time solutions  for $\eps=0.05$ 
are given by solid lines and have converged close to a stationary state.
In (a), this stationary profile is bounded between $[-1,1]$, where we
emphasize that $u$ in the left inset is 
still below 1 (dashed line in the inset), while in (b), the upper bound 1 is exceeded for 
$r$ less than about 0.4 (see left inset in (b)). Notice that in both (a) and (b),
the value for $u$ for $r>0.7$ is close to but visibly larger than $-1$, by an
amount that is consistent with the $O(\eps)$ correction predicted by the
asymptotic analysis (for (a) in \eqref{ostat}).}
\label{comparing_initial_condition}
\end{figure}

%%%%%%%%%%%%%%%%%%

The resolution to the above conundrum comes from the observation that
for a degenerate mobility, slowly evolving solutions can arise from
situations other than constant $\mu$ once
$|u|$ gets close to 1. To obtain an indication of how such
solutions evolve, we look at numerical solutions of the radially
symmetric version of \chefab{} on the domain $\Omega=\{(x,y); \,
r<1\}$, where $r=(x^2+y^2)^{1/2}$, starting with a tanh as initial
profile such that $ u_{\mathrm{init}}(r)<1$.  The spectral method we
used is briefly described in the appendix.  The numerical solution
at a later stage as shown in Fig.~\ref{comparing_initial_condition}
is positive for $r<0.5$ and negative for $r>0.5$.  Notice while
for $r>0.6$ the solution for $u$ levels out into a flat state that
is larger than $-1$ by an amount of $O(\eps)$, for $r<0.4$
the solution is much closer to $u=1$. Closer inspection shows that $u$ has a
maximum which approaches $u=1$, say at $r=r^*$. 
The maximum of $u$
may touch $u=1$ in either finite or infinite time. In either case,
the solution in $\Omega_+$ splits into two parts to the left and right of
$r^*$.  
The flux between the two parts is very small, and this suggests
that they are nearly isolated from each other.  In particular, they
do not have to be at the same chemical potential. Since we are only
interested in the phase field where it determines the evolution of
the interface, we cut off the part with $r<r^*$, and consider the
remaining part $r>r^*$ as a free boundary problem.

Returning to the general case of not necessarily radially symmetric
solutions, we introduce a free boundary $\Gamma$ 
near the interface inside $\Omega_+$, and cut off the parts of the 
solution further inside of $\Omega_+$. At $\Gamma$, we impose
\begin{equation}\label{chefb_freebcs}
u=1, \qquad \mathbf{n}_\Gamma \cdot \mathbf{j}=0, 
\qquad
\mathbf{n}_\Gamma \cdot \nabla u=0.
\end{equation}
\end{subequations}
Notice that in addition to $u=1$ and vanishing normal flux, a third
condition has been introduced at $\Gamma$. This is expected for
non-degenerate fourth order problems and permits a local expansion
satisfying \eqref{chefb_freebcs} that has the required number of two
degrees of freedom~\cite{KingB01}.  Indeed, expanding the solution
to \eqref{chefb} in a travelling wave frame local to $\Gamma$
with respect to the coordinate $\eta$ normal to $\Gamma$ gives
$u=1-a\eta^2+O(\eta^3)$, where $a$ and the position of the free boundary
implicit in the travelling wave transformation represent the two degrees of freedom.
 
Also observe that if $u>-1$ by $O(\eps)$ as suggested by the numerical
solution in Fig.~\ref{comparing_initial_condition}(a), then
$M(u)=O(\eps)$. Since $\mu=O(\eps)$, we expect a nonlinear
bulk flux of order $O(\eps^2)$ at the interface arising from
$\Omega_-$. This is the same order as the expected flux from surface
diffusion. Indeed, as shown below, both contributions are present in the leading
order sharp interface model~\eqref{one_side_porous_mediumd}.

Another scenario is conceivable if the mobility is changed to $|1-u^2|$.
Then, with an appropriate initial condition, we obtained numerical
results for the radially symmetric case which suggest a solution
that is not confined to $|u|<1$ and which in fact converges to the usual
stationary Cahn-Hilliard solution (considered, for example,
in \cite{Nieth95})
for which $\mu$ is
constant in $\Omega$, and $u$ is larger than one in most of $\Omega_+$.
These results are shown in Fig.~\ref{comparing_initial_condition}(b).
In this case, bulk fluxes from both $\Omega_+$ and $\Omega_-$ contribute to the leading order interface dynamics, 
see section~\ref{init_con_sil}.

%%%%%%%%%%%%%%%%%%%%%%%%%%%%%%%%%%%%%%%%%%%%%%%%%%%%%%%%%%%%%%%%%%%%%%%%%%%%%%%%%%%%%%%%%%%%%%%

\section{Radially symmetric stationary solution}\label{sec:axi}

By setting $u_\tau=0$ in \eqref{chefb} for a radially symmetric domain
$\Omega=\{(x,y); r<1\}$ and radially symmetric
$u=u(r)$, where $r=(x^2+y^2)^{1/2}$
and then integrating we obtain
\begin{subequations}\label{eqche}
\bl
\begin{align}
\frac{\eps^2}{r} \frac{\mathrm{d}}{\mathrm{d} r} \left( r \frac{\mathrm{d} u}{\mathrm{d} r} \right)+ \eta-2u(u^2-1) &=  0,
\label{ode_bvp_freebdyproblem}
\\
u'(1) &= 0,\label{newradbc} \\ 
u(r^*) = 1, \qquad u'(r^*) &= 0.
\label{free_bdy_condition}
\end{align}
\el
The point $r^*$ represents the location of the free boundary $\Gamma$ that needs to
be determined as part of the problem. The chemical potential $\eta$ is constant that needs to be determined by fixing the size of
the $\Omega_+$. This can be done by specifying
the $\int_\Omega u$, or, simpler, the 
position $r_0$ of the interface,
\bl
\begin{equation}
u(r_0) = 0.
\label{free_bdy_interface}
\end{equation}
\el
\end{subequations}
Note that if we do not consider a free boundary $\Gamma$ and 
impose $u'(0)=0$ instead of \eqref{free_bdy_condition}, 
then there exist exactly two solutions 
(which can be discerned by the sign of $u(0)$) as was shown
in \cite{Nieth95}.
We will now investigate \eqref{eqche} 
in the sharp interface limit $\eps\to 0$ using matched asymptotics.
There is one outer region away from the interface, and 
two inner layers, one located at the interface $r_0$ and one
located at $r^*$.

\subsection*{Outer region}
Inserting the ansatz 
\bl
\begin{align*}
u &= u_0 + \eps u_1 + \cdots, \qquad
\eta = \eta_0+ \eps \eta_1 +  \cdots,  
\end{align*}
\el
into \eqref{ode_bvp_freebdyproblem} and \eqref{newradbc} and
taking into account that the chemical potential $\eta$ is a constant quickly
reveals that $u_0$, $u_1$ and $u_2$ are also constants.  Their values
are fixed by standard matching, that is, they are equal
to the limits of the inner solutions as $\rho\to\infty$,
which therefore have to be bounded in this limit.

\subsection*{Inner layer about the interface}
To elucidate the asymptotic structure of the interface, 
we strain the coordinates about $r_0$ and write
\begin{equation}
\rho = \frac{r - r_0}{\eps},
\end{equation}
so that for $U(\rho)=u(r)$, and with the interface curvature $\kappa=1/r_0$, 
we have
\begin{equation}
\ti{U}'' + \eps \frac{\ti{U}'}{\kappa^{-1} + \eps \rho} + \eta -2 (\ti{U}^3-\ti{U}) = 0,
\qquad U(0)=0.
\end{equation} 
Expanding
\bl
$ 
\ti{U} = \ti{U}_0 + \eps \ti{U}_1 + \cdots,
$
\el
we have, to leading order,
\begin{equation}\label{u_0_sol}
\ti{U}_0''  -2 (\ti{U}_0^3-\ti{U}_0) = \eta_0,
\qquad U_0(0)=0.
\end{equation} 
To match with the outer and the solution near  $\Gamma$, $\ti{U}_0$ needs
to be bounded for $\rho\to\pm \infty$, which gives
\begin{equation}
\ti{U}_0 = - \tanh \rho, \quad \eta_0=0. 
\end{equation}
To $O(\eps)$ we have
\begin{equation}
\ti{U}_1''  -2 (3 \ti{U}_0^2-1)\ti{U}_1 = -\eta_1 - \kappa \ti{U}_0',
\quad
\ti{U}_1(0) =0,
\label{firsto_inner}
\end{equation}
for which the solution that is bounded as $\rho \rightarrow \infty$ is given by
\begin{eqnarray}
\ti{U}_1 &=& 
-\frac{1}{16} (\eta_1 + 2\kappa )
\text{sech}^2 \rho + 
\frac{1}{3} (3\eta_1 - 2\kappa)
\text{sech}^2 \rho \left( \frac{3 \rho}{8} + \frac{1}{4} \sinh 2\rho + \frac{1}{32}\sinh 4 \rho \right) \nonumber \\ 
&& + \frac{1}{8} (2 \kappa - \eta_1) + \frac{1}{48} (2\kappa - 3 \eta_1) (2 \cosh 2\rho - 5 \; \text{sech}^2 \rho). 
\label{u_1_soln}
\end{eqnarray}

\subsection*{Inner layer about $\Gamma$}
We centre the coordinates about the free boundary $r=r^*$ and write
\begin{equation}
z = \rho + \sigma,
\qquad
\sigma\equiv(r_0-r^*)/\eps.
\end{equation}
Substituting in the ansatz
$
\bar{U} = 1 + \eps \bar{U}_1 + \eps^2 \bar{U}_2+\ldots, 
$
we obtain, to $O(\eps)$, the problem
\begin{subequations}
\bl
\begin{align}
\bar{U}_1'' - 4 \bar{U}_1 &= -\eta_1, 
\label{fistoder_freebdy}
\\
\bar{U}_1(0) &= 0, \quad \bar{U}_1'(0) =0, 
\label{init_con_interior_firstord}
\end{align}
\el
\end{subequations}
with the solution
\begin{equation}
\bar{U}_1 = \frac{\eta_1}{4} \left( 1 - \cosh 2 z \right).
\label{v1}
\end{equation}

\subsection*{Matching}
\label{exponential_matching}

We first observe from \eqref{free_bdy_condition} that
the location of the free boundary $\Gamma$ in the inner coordinate
$\rho= -\sigma$ satisfies $U(-\sigma)=1$, $U'(-\sigma)=0$.
However, for $\eps\to 0$, we also have $U\to U_0=-\tanh(\rho)<1$.
To reconcile these conditions, we need to assume $\sigma\to \infty$
as $\eps \to 0$.
Matching of the inner expansions therefore involves 
exponential terms with large negative arguments $\rho$, or conversely for
large positive $z$, which we deal with in the spirit of Langer \cite{langer1983}, see also
\cite{korzec2008stationary}. The solution centred at the interface is
expanded at $\rho\to-\infty$ and the result written and re-expanded in
terms of $z=\rho+\sigma$. Notice that this change of variables can lead
to terms changing their order in $\eps$ if $\sigma$ has the appropriate magnitude.
The solution for the layer around the free boundary $\Gamma$ 
is directly expanded
in terms of $z\to\infty$ and then the terms are matched between the two expansions.

Expanding $U_0$ and $U_1$ for $\rho\to -\infty$ and substituting $\rho=z-\sigma$ gives
\begin{eqnarray}
\ti{U} &=& \Big( 1 - \underbrace{2 \rm  e^{-2\sigma} \rm e^{2z}}_{\mytag{A}{termA}} + O(\rm \rm e^{4 z}) \Big) + \eps \left\{ \underbrace{\frac{1}{24} (2\kappa - 3\eta_1)\rm e^{2\sigma} \rm e^{-2z}}_{\mytag{B}{termB}} +  \underbrace{\frac{1}{2} (\kappa - \eta_1)}_{\mytag{C}{termC}} \right. \nonumber \\ 
&& \left.+\underbrace{ \left[ \left( \frac{7\eta_1}{4} - \frac{11 \kappa}{6}\right) +\left( \frac{3 \eta_1}{2} -\kappa \right)(z-\sigma) \right] \rm e^{-2\sigma} \rm e^{2z}}_{\mytag{D}{termD}} + O(\rm e^{4z} ) \right\} \nonumber \\ 
&& + O(\eps^2). 
\label{asym_series_1}
\end{eqnarray}
The inner expansion for $\bar U$ 
at $z\to\infty$ is
\begin{equation}
\bar{U} = 1 + \underbrace{ \frac{\eps \eta_1}{4}}_{\mytag{E}{termE}} -  \underbrace{\frac{\eps \eta_1}{8} \rm e^{2z}}_{\mytag{F}{termF}}  - \underbrace{\frac{\eps \eta_1}{8}  \rm e^{-2z}}_{\mytag{G}{termG}} + O(\eps^2).  
\label{asym_series_2}
\end{equation}
Comparing terms in (\ref{asym_series_1}) and (\ref{asym_series_2})
of the same order of $\eps$ functional dependence with respect to $z$,
we notice first that the constant terms at $O(1)$ are already matched. Matching
$\eps$\ref{termC} and \ref{termE}, 
yields
\begin{equation}
\eta_1 = \frac{2}{3} \kappa.
\label{eta_1_eq}
\end{equation}
As a result, the term \ref{termB} is zero.
Matching term \ref{termA} and \ref{termF}, we arrive at 
the condition 
$2 \rm e^{-2\sigma} = {\eps\kappa}/{12}$,
which we solve for $\sigma$, giving
\bl
\begin{equation}
\sigma = \frac{1}{2}\log\left( \frac{24}{\eps\kappa}\right). 
\end{equation}
\el

We can now determine the outer solutions.
We note that in the more general, time
dependent situation,
the presence of a non-zero correction will give rise to a flux
at $O(\eps^2)$.
Using the limits of $U_0$ and 
$U_1$ as $\rho\to\infty$, we obtain
\begin{equation}\label{ostat}
u_0 = - 1 , \qquad
u_1 = \frac{\kappa}{6}.
\end{equation}

\subsection*{Higher corrections}
At this stage, it is obvious that the matching is not yet complete to $O(\eps)$,
as the terms in
\eqref{asym_series_2} and
\eqref{asym_series_1}, respectively,
$\eps$\ref{termD} and \ref{termG} 
are non-zero and lack counterparts in the other expansion.
This can be resolved by considering the next higher order solutions 
$\bar{U}_2 $ and $U_2$, which, in fact, will also be useful in section~\ref{sec:sid}.
We include $\eps^2\eta_2$ in the expansion for $\eta$,
and allow for corrections to $\sigma$
via the expansion
\begin{equation}\label{sigexp}
\sigma =  \frac{1}{2}\log\left( \frac{24}{\eps\kappa}\right)+\eps \sigma_1 + \cdots.
\end{equation}
The $O(\eps^2)$ problem at the interface is given by
\bl
\begin{align}
\ti{U}_2''  -2 (3 \ti{U}_0^2-1)\ti{U}_2 &= -\eta_2 - \kappa \ti{U}_1' + \rho \kappa^2 \ti{U}_0' + 6 \ti{U}_0 \ti{U}_1^2 \nonumber \\ 
&= -\eta_2 - \frac{\kappa^2}{6} \tanh^5 \rho - \rho \kappa^2 \text{sech}^2 \rho  - \frac{\kappa^2}{3} \tanh \rho \; \text{sech}^2 \rho, 
\label{second_ord_inner}
\end{align} together with 
\el
$U_2(0)=0$ and boundedness for $U_2$ as $\rho\to\infty$.  
The solution is
\bl
\begin{align}
\ti{U}_2 &= - \frac{\eta_2}{8} - \frac{\rho \kappa^2}{4} - \frac{1}{8} \cosh 2 \rho \left( \eta_2 + \frac{2}{3} \rho \kappa^2 \right) + \frac{1}{16} \text{sech}^2 \rho \left( 5 \eta_2 + \frac{23}{6} \rho \kappa^2 - 2 \rho^2 \kappa^2 \right) \nonumber \\
&\quad+ \frac{1}{4} \rho \kappa^2 \log \left( \frac{1}{2} \rm e^{\rho} \right) \text{sech}^2 \rho + \frac{\kappa^2}{8} \text{sech}^2 \rho \; \mathrm{Li}_2 (-\rm e^{2\rho}) 
\nonumber \\ 
&\quad - \frac{\kappa^2}{288} \sinh 2\rho \left( 1- 24 \log \cosh \rho \right) 
\nonumber \\ 
&\quad - \frac{\kappa^2}{96} \tanh \rho \left( 1- 24 \log \cosh \rho - \frac{8}{3} \text{sech}^2 \rho \right)
+  \frac{1}{16}\left(\frac{\pi^2}{6}  \kappa^2 - \eta_2 \right)
\text{sech}^2 \rho 
\notag\\
&\quad
+\left( \frac{\kappa^2}{36} ( 1+ 24 \log 2)  + \eta_2\right)
\text{sech}^2 \rho \left( \frac{3 \rho}{8} + \frac{1}{4} \sinh 2\rho +\frac{1}{32} \sinh 4 \rho \right),
\label{fullsol_u2}
\end{align}
\el
where $\mathrm{Li}_2 (x)$ is the dilogarithm function. 

For $\bar{U}_2(z)$ we have 
\begin{subequations}
\bl
\begin{align}
\bar{U}_2 '' - 4 \bar{U}_2 + \kappa \bar{U}_1' - 6 \bar{U}_1^2  + \eta_2&= 0,\\
\bar{U}_2(0) = 0, \quad \bar{U}_2'(0) &=0, 
\end{align}
\el
\end{subequations}
which has the solution
\bl
\begin{align}
\bar{U}_2 &=\left( \frac{\kappa}{12}\right)^2 (\cosh 4z + 3\rm e^{-2z} (1+ 4z) - 9 ) + \left(\frac{\kappa}{12}\right)^2 \rm e^{2z}
\notag
\\ & \qquad + \left(\frac{\kappa}{6} \right)^2 \rm e^{-2z} + \frac{\eta_2}{4} (1- \cosh 2z). 
\label{second_ord_free_bdy}
\end{align} 
\el

Expanding $U=\ti{U}_0+\eps\ti{U}_1+\eps^2U_2+\cdots$ for $\rho \rightarrow -\infty$,
substituting in $\rho=z-\sigma$ and using \eqref{sigexp} leads to 
\begin{eqnarray}
\ti{U} &=& 1 - \frac{\eps \kappa}{12} \rm e^{2 z} (1 - 2 \eps \sigma_1 ) 
+ \frac{1}{2} \left( \frac{\eps \kappa}{12}\right)^2 \rm e^{4z}  
+ \eps 
\left(\frac{\kappa}{6} - \frac{\eps \kappa^2}{36} \rm e^{2 z} \right) \nonumber \\ 
&&+ \eps^2 \left[ -\frac{1}{8} \eta_2 \left( \frac{24}{\eps \kappa} \right) (1+2 \eps \sigma_1)  \rm e^{-2z} + \left( \frac{\eta_2}{4} - \frac{\kappa^2}{16} \right)  \right]
+ O(\eps^{3}). 
\label{com_u1_u2}
\end{eqnarray}
Similarly, the expansion for $\bar{U}=\bar{U}_0+\eps\bar{U}_1+\eps^2U_2+\cdots$ 
as $z\to\infty$ is 
\bl
\begin{align}
\bar{U} &= 1 + \eps \frac{\kappa}{6} \left( 1- \cosh 2z \right)  \nonumber \\ 
&\quad + \eps^2 \left[ \frac{1}{2} 
\left(\frac{\kappa}{12}\right)^2 \rm e^{4z} + \frac{1}{2} 
\left(\frac{\kappa}{12}\right)^2 \rm e^{-4z} 
+ \left(\frac{\kappa}{12}\right)^2 ( 3\rm e^{-2z} (1+ 4z) - 9 )  \right. \nonumber \\ 
&\left. \;\;\;\ +  \left(\frac{\kappa}{12}\right)^2  \rm e^{2z} +  \left(\frac{\kappa}{6}\right)^2  \rm e^{-2z} + \frac{\eta_2}{4} (1-\cosh 2z) \right]. 
\label{second_order_freebdy}
\end{align} 
\el
Now, we can match the $\rm e^{-2z}$ at $O(\eps)$ and the  
$\rm e^{2z}$ at $O(\eps^2)$ terms, 
and arrive at, respectively, 
\bl
\begin{align}
\label{val_eta_2_sig_1}
\eta_2 &= \frac{\kappa^2}{36},
\qquad
\sigma_1 = \frac{3 \kappa}{16}.
\end{align}
\el
For completeness 
we note that the next order outer correction $u_2$ is again a constant equal
to the limit of $U_2$ as $\rho\to\infty$, with the value
$u_2 = {7\kappa^2}/{144}$.

\begin{figure}
\centering
	\subfloat{
		\includegraphics[width=0.50 \textwidth]{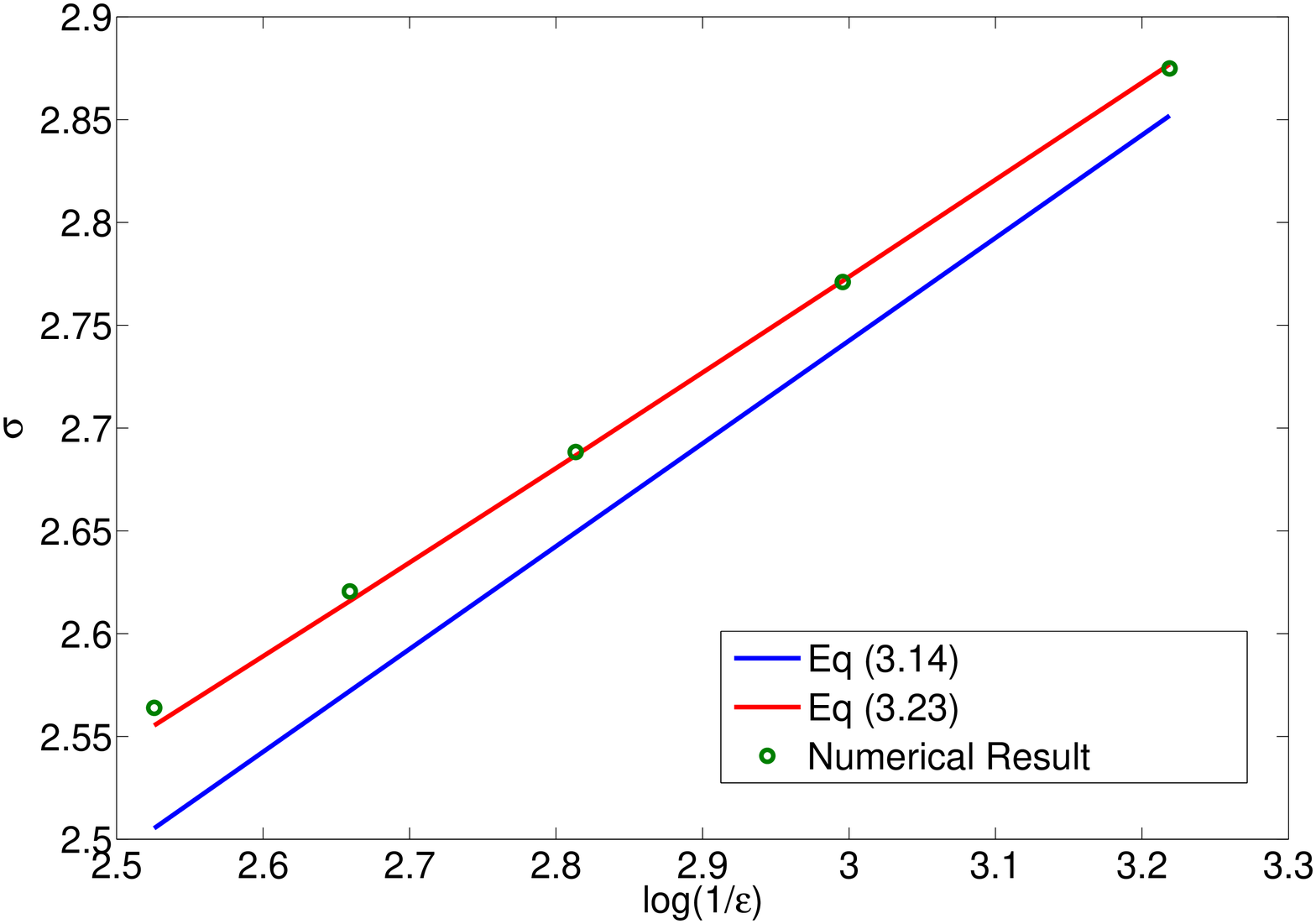}
	}
	\subfloat{
		\includegraphics[width=0.455 \textwidth]{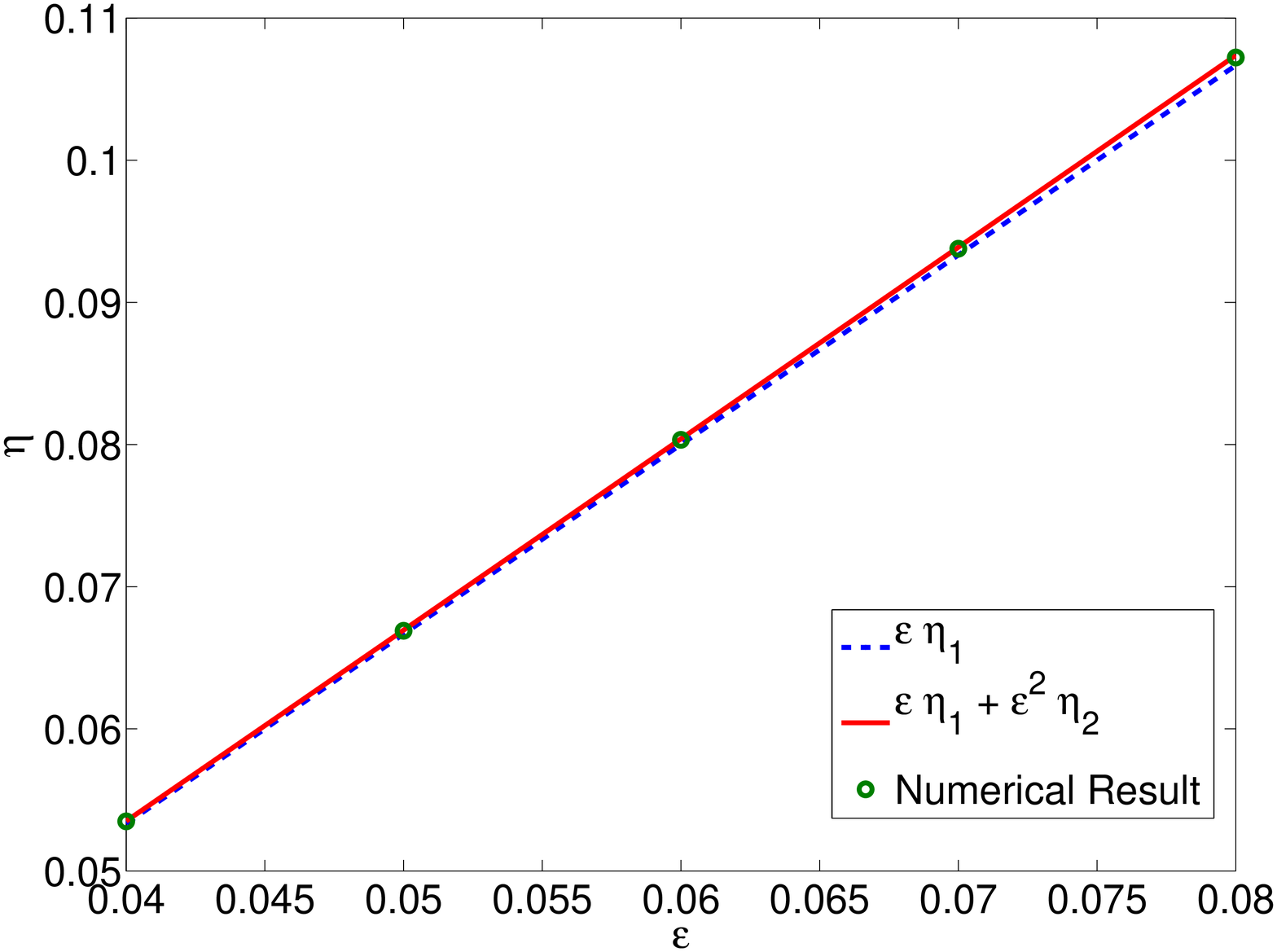}
	}
	\caption{Comparing the asymptotic and numerical results for (left) the position of the free boundary and (right) the chemical potential, for a range of $\eps$ and $r_0=1/2$.}
		\label{num_asyp_expo}
	\end{figure}

Figure \ref{num_asyp_expo} shows that the asymptotic results agree well
with the position of $\Gamma$ and the chemical potential obtained from
numerical solutions of the ODE free boundary problem (\ref{eqche}),
confirming the validity of the matched asymptotic results. The solutions
were obtained by a shooting method with fixed $\eta$ using the Matlab
package \textit{ode15s}, with $u(1)$ and \eqref{free_bdy_condition} as
the shooting parameter and condition.  The value of $\eta$ is adjusted
in an outer loop via the bisection method until $r_0=1/2$ is achieved to a
$10^{-10}$ accuracy.

%%%%%%%%%%%%%%%%%%%%%%%%%%%%%%%%%%%%%%%%%%%%%%%%%%%%%%%%%%%%%%%%%%%%%%%%%%%%%%

\section{Sharp Interface Dynamics}\label{sec:sid}
\subsection{Outer variables}

Motivated by the stationary state, we now consider the asymptotic
structure of the dynamical problem that arises for non-radially symmetric
interface geometries.  For the outer expansions, we will use
\bl
\begin{align*}
u & = u_0 + \eps u_1 + \eps^2 u_2 + \cdots, \quad %\\ \nonumber  
\mu  = \mu_0 + \eps \mu_1 + \eps^2 \mu_2 + \cdots, \quad %\\
\mathbf{j}=\mathbf{j}_0 + \eps \mathbf{j}_1 + \eps^2 \mathbf{j}_2 + \cdots.
\end{align*}
\el

\subsection{Inner variables}
As in \cite{pego1989front, gugenberger2008comparison}, we define
the local coordinates relative to the position of the interface
(parametrised by $s$), and write
\begin{equation} 
\mathbf{r}(s,r,\tau) = \mathbf{R}(s,\tau) + r \mathbf{n}(s,\tau), 
\end{equation} 
where $\mathbf{R}$, the position of the interface $\zeta$, is defined by
\begin{equation}\label{zetapos}
u(\mathbf{R},t)=0,
\end{equation}
and $\mathbf{t} = \partial \mathbf{R}/\partial s$ is the unit tangent vector, and $\mathbf{n}$ is the unit outward normal. From
the Serret-Frenet formulae in 2D we have that 
$\kappa \mathbf{t} = \partial \mathbf{n}/\partial s$, thus
\begin{equation}
\frac{\partial \mathbf{r}}{\partial r} =   \mathbf{n}(s), 
\qquad
\frac{\partial \mathbf{r}}{\partial s} =   (1+r\kappa) \mathbf{t}(s),
\end{equation} 
where $\mathbf{t}(s)$ is the unit tangent vector to the interface, and $\kappa$ is the curvature. We adopt the convention that the curvature is positively defined if the osculating circle lies inside $\Omega_+$. The gradient operator in these curvilinear coordinates reads 
\begin{equation}
\nabla = \mathbf{n} \partial_r + \frac{1}{1+r\kappa} \mathbf{t} \; \partial_s,  
\label{grad_in_local}
\end{equation} 
and the divergence operator of a vector field $\mathbf{A} \equiv A_r \mathbf{n}+ A_s\mathbf{t}$ reads
\begin{equation}
\nabla \cdot \mathbf{A} =  \frac{1}{1+r\kappa} \left[ \partial_{r}\Big( (1+r \kappa) A_{n} \Big) +  \partial_{s} \left(\frac{1}{1+r\kappa} A_{s} \right)   \right]. 
\label{div_in_local}
\end{equation}

We let $s$ and $\rho=r/\eps$ be the inner coordinates at the interface,
and  let $U(\rho,s,\tau)$, $\eta(\rho,s,\tau)$ and $\mathbf{J}(\rho,s,\tau)$
denote the order parameter, chemical potential 
and flux written in these coordinates, respectively. 
In inner coordinates, the combination of the first two equations, in
\eqref{chereb}, and \eqref{zetapos}, become
\begin{subequations}\label{chin}
\begin{align} \label{china}
\eps^2 \partial_\tau \ti{U} - \eps v_n \partial_\rho \ti{U} 
&= \nabla \cdot \left( M(\ti{U}) \nabla \ti{\eta} \right), \\
\eta &= - \eps^2 \nabla^2 U + f'(U),\\
U(0) &=0,
\end{align} 
with $v_n=\bf{R}_\tau\cdot n$. Using equations (\ref{grad_in_local}) and (\ref{div_in_local}), we obtain 
\begin{eqnarray}
\nabla \cdot \left( M(\ti{U}) \nabla \right) &=& \eps^{-2} \partial_{\rho} M(\ti{U}_0) \partial_{\rho}  \nonumber\\
&&+ \eps^{-1} \Bigg\{ \partial_{\rho} \Big(\kappa \rho M(\ti{U}_0) + M'(\ti{U}_0) \ti{U}_1 \Big) \partial_{\rho}  - \kappa \rho \; \partial_\rho M(\ti{U}_0) \partial_\rho \Bigg\} \nonumber \\ 
&&+ \bigg\{ \kappa^2 \rho^2 \partial_\rho M(\ti{U}_0) \partial_{\rho} - \kappa \rho \partial_\rho \Big(\kappa \rho M(\ti{U}_0) + M'(\ti{U}_0) \ti{U}_1 \Big) \partial_\rho \nonumber \\  
&& + \partial_\rho \Big(\kappa \rho M'(\ti{U}_0) \ti{U}_1 + \frac{1}{2} M''(\ti{U}_0) \ti{U}_1^2 + M'(\ti{U}_0) \ti{U}_2 \Big) \partial_\rho  \nonumber \\ 
&&+ \partial_s M(\ti{U}_0) \partial_s  \bigg\}+O(\eps).
\label{chinb}
\end{eqnarray}
\end{subequations}
Notice that the corresponding expression for $\nabla^2$ can be easily obtained from this
by setting $M\equiv 1$.

Taking only the first equation in \eqref{chereb} we have
\begin{equation} 
\eps^2 \partial_\tau \ti{U} - \eps v_n \partial_\rho \ti{U} = 
\frac{1}{1+\eps\rho\kappa} \left[\eps^{-1} \partial_{\rho}\Big( (1+\eps\rho\kappa) J_{n} \Big) +  \partial_{s} \left(\frac{1}{1+\eps\rho\kappa} J_{s} \right)   \right]. 
\end{equation} 
In inner coordinates, we will only need to know the normal component
$J_n=\mathbf{n}\cdot \mathbf{J}$ of the flux explicitly in terms of
the order parameter and chemical potential. It is given by
\bl
\begin{align}
J_n
&= \frac{M(\ti{U})}{\eps} \partial_\rho \ti{\eta} \nonumber \\ 
&= 
\eps^{-1} M(\ti{U}_0) \partial_{\rho} \ti{\eta}_0 
+M'(U_0) U_1 \partial_\rho \eta_0+M(\ti{U}_0) \partial_{\rho} \ti{\eta}_1 
\notag\\
&\quad+ \eps \left( M(\ti{U}_0) \partial_{\rho}\ti{\eta}_2 
		+  M'(\ti{U}_0) \ti{U}_1 \partial_\rho \ti{\eta}_1 
			+ M'(\ti{U}_{0}) \ti{U}_2 \partial_{\rho} \ti{\eta}_0
				+\frac{1}{2} M''(\ti{U}_0) \ti{U}_1^2 \partial_{\rho} \eta_0 \right) \nonumber \\ 
&\quad+ 
\eps^2 \left[ M(\ti{U}_0) \partial_{\rho} \ti{\eta}_3 + M'(\ti{U}_0) \ti{U}_1 \partial_{\rho}
	 \ti{\eta}_2  + \left( M'(\ti{U}_0) \ti{U}_2 + \frac{1}{2} M''(\ti{U}_0) \ti{U}_1^2 
		 \right)\partial_{\rho} \ti{\eta}_1 \right. \nonumber \\ 
&	\qquad\qquad	+ \left. \left( M'(\ti{U}_0) \ti{U}_3 + M''(\ti{U}_0) \ti{U}_1\ti{U}_2 
		+ \frac{1}{6} M'''(\ti{U}_0) \ti{U}_1^3 \right)\partial_{\rho} 
		\ti{\eta}_0  \right] 
+ O(\eps^3),
\label{inner-flux}
\end{align} 
\el
which also motivates our ansatz for the expansion for $\mathbf{J}$,
given the obvious ansatz for the other variables,
\bl
\begin{align*}
U & = U_0 + \eps U_1 + \eps^2 U_2 + \cdots, \quad %\\ \nonumber  
\eta  = \eta_0 + \eps \eta_1 + \eps^2 \eta_2 + \cdots,  \\ %\quad %\\
\mathbf{J}&=\eps^{-1}\mathbf{J}_{-1} 
+\mathbf{J}_0 + \eps \mathbf{J}_1 + \eps^2 \mathbf{J}_2 + \cdots.
\end{align*}
\el

Moreover,  we introduce $z = \rho + \sigma(s,t)$ as the coordinate for the inner layer
about the the free boundary $\Gamma$, so that the order parameter, chemical potential and flux
in these variables are given by $\bar{U}(z,s,\tau)$, 
$\bar{\eta}(z,s,\tau)$ and $\bar{\mathbf{J}}(z,s,\tau)$, respectively, with expansions
\bl
\begin{align*}
\bar{U} & = \bar{U}_0 + \eps \bar{U}_1 + \eps^2 \bar{U}_2 + \cdots, \quad %\\ \nonumber  
\bar{\eta}  = \bar{\eta}_0 + \eps \bar{\eta}_1 + \bar{\eps}^2 \bar{\eta_2} + \cdots,\\
\bar{\mathbf{J}}&=
\eps^{-1}\bar{\mathbf{J}}_{-1}+
\bar{\mathbf{J}}_0 + \eps \bar{\mathbf{J}}_1 + \eps^2 \bar{\mathbf{J}}_2 + \cdots.
\end{align*}
\el

Notice that the location where the two inner layers are centred
depends on $\eps$ and therefore, in principle, $\sigma$ and also $R$
need to be expanded in terms of $\eps$ as well. However, we are only
interested in the leading order interface motion, so to keep the notation
simple, we do not distinguish between $\sigma$ and $R$ and their leading
order contributions.  We now solve and match the outer and inner problems
order by order.

\subsection{Matching}\label{subsec:ord_sol}
\subsubsection*{Leading order}

For the outer problem, we obtain to leading order
\begin{equation} 
\nabla \cdot \mathbf{j}_0 = 0, \quad
\mathbf{j}_0 = M(u_0) \nabla \mu_0 , \quad
 \mu_0 = f'(u_0). 
\label{eliptic_pde} 
\end{equation} 
The requisite boundary conditions are $\nabla_n u_0 = 0 $, and 
$\mathbf{n}\cdot \mathbf{j}_0= 0 $ on $\partial \Omega$. 
We have
\begin{equation}
u_0 = - 1, \qquad \mu_0 = 0. 
\end{equation} 

The leading order expansion about the interface reads,
\begin{equation} 
M(\ti{U}_0 )\partial_{\rho} \ti{\eta}_0  = a_1(s,\tau),
\quad
f'(\ti{U}_0) - \partial_{\rho\rho} \ti{U}_0 = \ti{\eta}_0.\label{U0}
\end{equation}
From the matching conditions, we require $U_0$ to be bounded for
$\rho\to\pm \infty$. In fact, $U(\rho\to-\infty)=-1$, giving $\eta_0\to
0$. This implies $a_1=0$, therefore also $\eta_0=0$, which we note
matches with $\mu_0$.  Moreover, from \eqref{U0}$_2$
and from \eqref{inner-flux} we have
\begin{equation}
\ti{U}_0 = -\tanh \rho,
\qquad 
J_{n,-1} = 0.
\end{equation} 
The leading order approximation of the order parameter 
in the coordinates of the inner layer at $\Gamma$ 
is easily found to be $\bar{\mathbf{U}}_0=1$, 
and also for the chemical potential $\bar{\eta}_0=0$,
and the normal component of the flux $\bar{J}_{n,-1} = 0$.

\subsubsection*{O($\eps$) correction}

The first two parts of the  outer correction problem for \eqref{chereb}
are automatically satisfied, since $\mu_0 = 0$ and $M(u_0) = 0$, by  
\begin{equation}\label{j1sol}
\mathbf{j}_1 = 0.
\end{equation}
The last part requires
\begin{equation}
\mu_1 = f''(u_0) u_1 = 4 u_1. 
\label{eq_chem_pot}
\end{equation} 
From \eqref{chin}, and noting that $\ti{\eta}_0 = 0$, we have
\begin{equation}
\partial_{\rho} \left( M(\ti{U}_0) \partial_{\rho} \ti{\eta}_1 \right)  = 0, 
\quad 
\ti{\eta}_1 = - \partial_{\rho\rho} \ti{U}_{1} - \kappa \partial_\rho \ti{U}_0 
+ f''(\ti{U}_0)\ti{U}_1,
\quad
U_1(0)=0,
\label{mu_const_o_eps}
\end{equation} 
thus $M(\ti{U}_0) \partial_{\rho} \ti{\eta}_1 = J_{n,0}$ is
constant in $\rho$. Since $J_{n,0}$ has to match with $j_0$, it
is zero.  Therefore, $\eta_1=\eta_1(s,t)$ does not depend on $\rho$.
Now (\ref{mu_const_o_eps})$_2$ and (\ref{mu_const_o_eps})$_3$ represent
the same problem as (\ref{firsto_inner}).  As such, the solution
$\ti{U}_1(\rho,s,\tau)$ that is bounded as $\rho \rightarrow \infty$ can be read
off (\ref{u_1_soln}).

The $O(\eps)$ problem for the inner layer at $\Gamma$ becomes 
\begin{equation}
\bar{\eta}_1 = - \partial_{zz} \bar{U}_{1} + 4 \bar{U}_1,  
\end{equation}
with $\bar{\eta}_1$ that does not depend on $z$,
supplemented with the conditions $\bar{U}_1(z,0,\tau) = 1$,
$\bar{U}_{1z}(z,0,\tau) = 0$. This equation is the same as the $O(\eps)$
equation for the stationary state about the free boundary, and the
solution is given by (\ref{v1}).
The inner layers about $\Gamma$ and about the
interface can be matched, as outlined in 
section~\ref{exponential_matching}, to obtain
\begin{equation}\label{eta1sol}
\bar\eta_1=\eta_1 = \frac{2}{3} \kappa. 
\end{equation}

\subsubsection*{O($\eps^2$) correction}

Combining the first two equations in \eqref{chereb}
and expanding to $O(\eps^2)$ yields
\begin{equation}
\nabla \cdot \left( M'(u_0) u_1 \nabla \mu_1 \right)= 0. 
\end{equation} 
In view of the discontinuous derivative of $M$ at $u=u_0=-1$, we
remark that here and in the following we will use the convention
that $M'(\pm 1)$ denotes the one-sided limit for $|u|\to 1^-$, in particular that
$M'(-1)=2$, and likewise for higher derivatives.
Equation (\ref{eq_chem_pot}) provides a relation between $\mu_1$ and $u_1$. 
Thus, we have 
\begin{equation}
\nabla \cdot \left(\mu_1 \nabla \mu_1 \right) =0
\label{porous_medium}
\end{equation}
with the boundary condition
$\nabla_n \mu_1 = 0$ on $\partial \Omega$,
and, from matching $\mu_1$ with $\eta_1$ (given in \eqref{eta1sol})
at the interface, 
\begin{equation}\label{mu1bc}
\mu_1 = \frac{2}{3} \kappa. 
\end{equation}
Expanding the second equation in \eqref{chereb} to $O(\eps^2)$ 
also gives us an
expression for the normal flux 
\begin{equation}\label{j2sol}
\mathbf{n}\cdot \mathbf{j}_2 = u_1 M'(u_0) \nabla_n \mu_1 = \frac{1}{2} \mu_1 \nabla_n \mu_1,
\end{equation} 
which is not in general zero.

\subsubsection*{Inner expansion about the interface}

From the $O(1)$ terms in \eqref{chin}, we obtain 
\begin{equation} 
\partial_\rho 
\left( M(\ti{U}_0) \partial_{\rho} \ti{\eta}_2 \right)=0. 
\label{eq_order_2} 
\end{equation}
Thus, $M(\ti{U}_0) \partial_{\rho} \ti{\eta}_2$ is constant
in $\rho$ and since we can identify this expression via \eqref{inner-flux}
as $J_{n,1}$, which has to match with $\mathbf{n}\cdot \mathbf{j}_1=0$. 
Therefore we can deduce that 
\begin{equation}
J_{n,1}=M(\ti{U}_0) \partial_{\rho} \ti{\eta}_2=0,
\end{equation}
and $\eta_2$ is independent
of $\rho$. The solution for $\eta_2$ is found in essentially the
same way as in Section~\ref{sec:axi}, see 
\eqref{sigexp} -- \eqref{val_eta_2_sig_1}, thus
\begin{equation}\label{eta2sol3} 
\eta_2(s,\tau)=\frac{\kappa^2}{36}.
\end{equation}

\subsubsection*{O($\eps^3$) correction}

Noting that $\ti{\eta}_0$, $\ti{\eta}_1$ and $\ti{\eta}_2$ are
independent of $\rho$, the $O(\eps)$ terms in 
\eqref{chin} yield
\begin{eqnarray} 
-v_n \partial_\rho \ti{U}_0 
&=&  \partial_\rho M(\ti{U}_0) \partial_\rho \ti{\eta}_3 +   \frac{2}{3}M(\ti{U}_0) \partial_{ss} \kappa.
\label{third_order} 
\end{eqnarray}
Integrating equation (\ref{third_order}) from $-\infty$ to $\infty$, 
we arrive at 
\begin{equation}\label{vnjump}
v_n  =\frac{1}{2} [M(\ti{U}_0) \partial_\rho \ti{\eta}_3 ]_{-\infty}^{\infty} + \frac{2}{3}  \partial_{ss} \kappa.  
\end{equation}
From \eqref{inner-flux}, we can identify the term in the bracket
as
\begin{equation}\label{Jn2}
J_{n,2}=M(\ti{U}_0) \partial_\rho \ti{\eta}_3.
\end{equation}
At $\rho\to-\infty$, we need to match $\eta_3$ and $J_{n,2}$ 
with the solution
for $\bar{\eta}_3$ and $\mathbf{n}\cdot\bar{\mathbf{J}}_2$ 
in the inner layer at $\Gamma$, which in the former case 
is a function independent of $z$, and in the latter is just 
zero. Thus, $\eta_3$ is matched to
a constant for $\rho\to-\infty$, and  $J_{n,2}$ is matched to zero, thus
\begin{equation}\label{J2minf}
\lim_{\rho \rightarrow -\infty} M(\ti{U}_0) \partial_{\rho} \ti{\eta}_3= 
\lim_{\rho \rightarrow -\infty}J_{n,2}=0. 
\end{equation}

We next consider the contribution from $J_{n,2}$ as $\rho\to\infty$. It
is tempting to use \eqref{Jn2} to argue that, since $M(U_0)\to0$
exponentially fast, $J_{n,2}$ also has to tend to zero.  
Then, however, $J_{n,2}$ cannot be be matched with $\mathbf{n} \cdot
\mathbf{j}_2$, as we cannot simply set the latter to zero: The bulk
equation \eqref{porous_medium} has already got a boundary condition
at $\zeta$, namely \eqref{mu1bc}, and setting $\mathbf{n} \cdot
\mathbf{j}_2=0$ would impose too many conditions there.
We therefore drop the idea that $J_{n,2}\to 0$ as $\rho\to\infty$ and
match the normal fluxes,
\bl
\begin{align}
\lim_{\rho \rightarrow \infty} J_{n,2} &= 
\left. \mathbf{n} \cdot \mathbf{j}_2\right|_\zeta, \label{Jj} 
\end{align}
\el
Keeping in mind that non-trivial solutions for $\mu_1$ will arise from
\eqref{porous_medium}, \eqref{mu1bc} and $\nabla_n \mu_1 = 0$ at $\partial
\Omega$, we expect that $J_{n,2}$ will not, in general
be zero because of \eqref{j2sol} and \eqref{Jj}.  
Substituting \eqref{Jn2}
and \eqref{j2sol} into the left and right hand sides of \eqref{Jj}, respectively, we obtain
\bl
\begin{align}
\lim_{\rho \rightarrow \infty} M(\ti{U}_0) \partial_{\rho} \ti{\eta}_3&=
\frac{1}{2} \mu_1 \nabla_n \mu_1 |_{\zeta}, \label{J2pinf}
\end{align}
\el
so that now the boundary terms in \eqref{vnjump} have been determined
in terms of $\mu_1$.
Now, however, we have to accept that in general there will be 
exponential growth in $\eta_3$ as $\rho\to\infty$: if the left hand side of \eqref{J2pinf} 
is nonzero, and $M(U_0)\to 0$ exponentially fast as $\rho\to\infty$, then
$\eta_3$ has to grow exponentially.
In fact, if we solve
\eqref{Jn2} for $\eta_3$, and eliminate $J_{n,2}$ via \eqref{Jj} and \eqref{j2sol},  
we obtain the solution
\begin{equation}\label{eta3}
\eta_3=\frac{\mu_1 \nabla_n \mu_1 |_{\zeta}}{16} 
\,\left(\rm e^{2\rho} + 2{\rho}\right) + \eta_3^0,
\end{equation}
where $\eta_3^0$ is an integration constant.
The term proportional $\rm e^{2\rho}$ is the exponentially
growing term and it does not appear to be matchable to the
outer solution.  We will resolve this issue in a separate section, by
introducing another inner layer, and for now continue with analysing
the sharp interface model, which in summary is given by
\begin{subequations}
\label{one_side_porous_medium}
\bl
\begin{align}
\nabla \cdot (\mu_1 \nabla \mu_1) &= 0, \quad \text{in } \Omega_+,  \\ 
\mu_1 &= \frac{2}{3} \kappa, \quad \text{on }  \zeta,  \\ 
\nabla_n \mu_1 &= 0 , \quad \text{on }  \partial \Omega_{\mathrm{ext}},\\
v_n &= \frac{2}{3}  \partial_{ss}  \kappa + \frac{1}{4} \mu_1 \nabla_n \mu_1\quad \text{on }  \zeta. \label{one_side_porous_mediumd} 
\end{align}
\el
\end{subequations}

%%%%%%%%%%%%%%%%%%%%%%%%%%%%%%%%%%%%%%%%%%%%%%%%%%%%%%%%%%%%%%%%%%%%%%%%%

\subsection{Additional inner layer}

The exponential growth of $\eta_3$ at $\rho\to\infty$ is a direct
consequence of the exponential decay of $M(U_0)$ to 0 as $U_0$
approaches $-1$ exponentially fast. Notice, however, that the inner
solution including the correction terms does not decay to $-1$,
because $U_1(\rho\to\infty)>0$,
so that 
\[
M(U_0+\eps U_1+\cdots)=
M(U_0)+\eps M'(U_0) U_1+\cdots
\]
approaches a non-zero $O(\eps)$ value as $\rho\to\infty$. We 
need to ensure that the correction $\eps M'(U_0) U_1$ to
$M(U_0)$ enters into the calculation of
the chemical potential as soon as $\rho$ is in the range where $M(U_0)$
and $ \eps M'(U_0) U_1$ have the same order of magnitude. This happens
when $U_0+1=O(\eps)$, i.e.\ when $\rho\sim -(1/2)\ln\eps$. We therefore
introduce another layer via
\bl
\begin{align*}
\rho&=\frac{1}{2}\ln\left(\frac1\eps\right)+y,
\quad
\hat{U}(y)=U(\rho),\quad
\hat{\eta}(y)=\eta(\rho),\quad
\hat {\mathbf{J}}(y)=\mathbf{J}(\rho).
\end{align*}
\el
Notice the similarity with the change of variables at $\Gamma$. Indeed,
the solution in the new layer will have exponential terms in the
expansion at $y\to-\infty$ that need to be matched with the
expansion at the interface
$\rho\to\infty$.
In terms of the new variables, the Cahn--Hilliard equation becomes
\bl
\begin{align} 
\eps^2 \partial_\tau \hat{U} - \eps v_n \partial_y \hat{U} 
&= \nabla \cdot \left( M(\hat{U}) \nabla \hat{\eta} \right),\\
\hat \eta &=
-\partial_{yy} \hat U
-\frac{\eps\kappa}{1+\eps\kappa\left(y-\frac12\ln\eps\right)}\partial_y \hat U 
\notag\\ &\quad 
-\frac{\eps^2}{1+\eps\kappa\left(y-\frac12\ln\eps\right)}
\partial_s\left(\frac{\partial_s \hat{U}}{1+\eps\kappa\left(y-
\frac12\ln\eps\right)}\right) + f'(\hat U).
\end{align} 
\el
We expand
\bl
\begin{align*}
\hat{U} & = -1 + \eps \hat{U}_1 + \eps^2 \hat{U}_2 + \cdots, \quad 
\hat{\eta}  = \eps \hat{\eta}_1 + \hat{\eps}^2 \hat{\eta_2} + \cdots,\nonumber\\
\hat{\mathbf{J}}&=
\hat{\mathbf{J}}_0 + \eps \hat{\mathbf{J}}_1 + \eps^2 \hat{\mathbf{J}}_2 + \cdots,
\end{align*}
\el
where we have tacitly anticipated that $\hat\eta_0=0$, $\hat{\mathbf{J}}_{-1}=0$.
Inserting these gives
\bl
\begin{align}
\nabla \cdot \left( M(\hat{U}) \nabla \hat{\eta} \right) &= 
 \partial_{y} \left[M'(-1) \hat{U}_1\partial_{y}\hat{\eta}_1\right]  
+\eps \partial_{y} \left[M'(-1) \hat{U}_1\partial_{y}\hat{\eta}_2\right]  
+O(\eps^2).
\end{align}
\el
The normal flux $\hat{J}_n=\mathbf{n}\cdot\hat{\mathbf{J}}$ is given by
\bl
\begin{align}
\hat{J}_n
&= \frac{M(\ti{U})}{\eps} \partial_\rho \ti{\eta} 
= \left[M'(-1)\hat{U}_1+
\eps \left(\left(M''(-1)/2\right) \hat{U}_1^2 + M'(-1)\hat{U}_2 \right) +
O(\eps^2)\right]
\nonumber\\
&\hspace*{30ex}
\times \left[\eps \partial_y 
\hat{\eta}_1+\eps^2
\partial_y\hat{\eta}_2+O(\eps^3)\right].\label{hatJn}
\end{align}
\el
Comparison with the ansatz for the expansion of $\hat{\mathbf J}$ immediately implies 
$\hat{J}_{n,0}=0$.

\subsection*{Leading order problem}

To leading order, we have
\bl
\begin{align}
-\partial_y\left[M'(-1) \hat{U}_1\partial_y \hat{\eta}_1 \right] &=0,
\qquad 
-\partial_{yy} \hat{U}_1+f''(-1) \hat{U}_1 = \hat{\eta}_1.
\end{align}
\el
Integrating the first of these once, we obtain that the expression in square bracket has
to be a constant in $y$. From \eqref{hatJn}, we see this is the term
$\hat{J}_{n,1}$ in the normal flux, which has to match to ${J}_{n,1}$
and $\mathbf{n}\cdot{\mathbf{j}}_1$ in the interface layer and the
outer problem, respectively. Thus
$\hat{J}_{n,1}=0$.
Therefore, the contribution $\hat{\eta}_1$ is also a constant that needs to match
to the same value $\kappa/6$ towards the outer and the interface layer, \emph{i.e.}\ for 
$\hat y\to \pm \infty$, so that we have
\bl
\begin{align}
\hat \eta_1 &= \frac23 \kappa,
\qquad
\hat{U}_1 = c_1  \mathrm{e}^{-2y}+ c_2 \mathrm{e}^{2y}+\frac16 \kappa.
\end{align}
\el
Matching this to the constant outer $u_1=\kappa/6$, obtained from
\eqref{eq_chem_pot} and \eqref{eta1sol}, forces $c_2=0$. We next expand
$U_0$ at $\rho\to\infty$,
\begin{equation}
U_0 = -1 + 2 \rm{e}^{-2\rho} + O( e^{-4\rho}).
\end{equation} 
The second term accrues a factor of $\eps$ upon passing to $y$-variables,
and thus has to match with the exponential term in
$\eps \hat U_1 $,
giving $c_1=2$ and 
\begin{equation}
\hat U_1 = 2 \mathrm{e}^{-2y} + \frac16\kappa.
\end{equation}

\subsection*{First correction problem}

To next order, we obtain
\begin{subequations}
\bl
\begin{align}
-\partial_y\left[M'(-1) \hat{U}_1\partial_y \hat{\eta}_2 \right] &=0,\label{nohata} \\
-\partial_{yy} \hat{U}_2-\kappa\partial_y\hat{U}_1
+f''(-1) \hat{U}_2+f'''(-1)\hat U_1 &= \hat{\eta}_2,\label{nohatb}\\
\hat{J}_{n,2}&= M'(-1) \hat{U}_1\partial_y\hat \eta_2.
\label{nohatc}
\end{align}
\el
\end{subequations}
From \eqref{nohata} and \eqref{nohatc}, and matching the flux contribution 
$\hat{J}_{n,2}$ to the outer $\mathbf{n}\cdot\mathbf{j}_2$,  we obtain
\bl
\begin{align}
M'(-1) \hat{U}_1 \partial_y \hat{\eta}_2 
= \frac{1}{2}\left.\mu_1\nabla_n\mu_1\right|_\zeta, 
\end{align}
\el
which in turn has the solution
\begin{equation}\label{heta2}
\hat \eta_2=\frac{\left.\mu_1\nabla_n\mu_1\right|_\zeta}{\kappa M'(-1)}\ln\left(\frac\kappa{12} \mathrm{e}^{2y}+1\right)+\frac{\kappa^2}{36}.
\end{equation}
The integration constant has been fixed by matching $\hat\eta_2$
for $y\to-\infty$ with the interface solution $\eta_2$, see \eqref{eta2sol3}. We now need
to check if the exponential term in \eqref{heta2} matches with the
exponential term in \eqref{eta3}.  Expanding at $y\to-\infty$ is
trivial, and then substituting in $y=\rho+\ln\eps/2$ gives
\begin{equation}
\hat \eta_2=\frac{\eps}{8M'(-1)}\left.\mu_1\nabla_n\mu_1\right|_\zeta 
\rm e^{2\rho}+\frac{\kappa^2}{36}.
\end{equation}
Thus, $\eps^2\hat\eta_2$ contains a term proportional to $\eps^3 \mathrm{e}^{2y}$ term that is identical to the $\eps^3 \mathrm{e}^{2y}$ term that
appears in $\eps^3 \eta_3$, see \eqref{eta3}. Thus, we have 
resolved the issue with the exponentially
growing term (for $\rho\to\infty$) in the correction to
the chemical potential in the interface layer expansion.

%%%%%%%%%%%%%%%%%%%%%%%%%%%%%%%%%%%%%%%%%%%%%%%%%%%%%%%%%%%%%%%%%%%%%%%%%

\subsection{Linear stability analysis}

\begin{table}
\centering
\begin{tabular}{| l | l | l | l | l | l | l | l |}
\hline 
 $\eps$ & 0.01 &  0.005  & 0.003 & 0.002 & 0.001  & \bf Eq (\ref{solid_diffusion_one_side_decay_rate}) & \bf Eq (\ref{pure_solid_diffusion_decay_rate}) \\ \hline
 $\lambda_{m=2}$ & $-$133.2  & $-$133.8  & $-$136.0 & $-$136.3 & $-$137.0 & \bf $-$137.4 
& \bf $-$128 \\ 
 \hline
\end{tabular}
\caption{\label{diffuse_sharp_deg1}
Relaxation rates obtained from the linearised phase field model
(\ref{lin_stab_pde_deg_mobility}) are shown for different values of $\eps$
in the first five columns, and compared to the eigenvalues obtained
for linearised sharp interface models for pure surface diffusion
\eqref{pure_solid_diffusion_decay_rate} and the porous medium type
model \eqref{solid_diffusion_one_side_decay_rate} in the next-to-last
and the last column, respectively, with $\mathfrak{M} = 2/3$. }
\end{table}

Besides the usual surface diffusion term, equation
(\ref{one_side_porous_medium}) contains an additional normal flux
term which is nonlocal. In cases where there are multiple regions
of $u$ close to 1, the
nonlocal term couples the interfaces of these regions with each other
and drive coarsening where the larger regions grow at the expense
of smaller ones.
This is not expected for pure surface diffusion.
Even for a single convex domain that is slightly perturbed
from its radially symmetric state, the effect on the relaxation dynamics
is noticeable, as we now explore.

To compare the sharp interface model with the phase field model, we consider the relaxation of an azimuthal perturbation to a radially symmetric stationary state with curvature $\kappa = 1/r_0$. For azimuthal perturbations proportional to $\cos m\theta$, the pure surface diffusion model $v_n = \mathfrak{M} \partial_{ss} \kappa$ predicts an exponential decay rate 
\begin{equation}
\lambda = - \mathfrak{M} \frac{m^2(m^2-1)}{r_0^4}.  
\label{pure_solid_diffusion_decay_rate}
\end{equation} 
In contrast, the decay rate in the porous medium model, Equation (\ref{one_side_porous_medium}), is given by
\begin{equation}
\lambda = - \frac{2}{3} \frac{m^2(m^2-1)}{r_0^4} - \frac{1}{9} \frac{m(m^2-1)}{r_0^4} \tanh (m \log r_0^{-1}).
\label{solid_diffusion_one_side_decay_rate}
\end{equation} 
In the diffuse interface model, the perturbation $v_1(r,t) \cos m\theta$ satisfies
\begin{eqnarray}
v_{1t} &=& \frac{1}{r} \frac{\partial}{\partial r} \left( r M(v_0) \frac{\partial \mathfrak{m}_1}{\partial r} \right) - \frac{m^2}{r^2} M(v_0) \mathfrak{m}_1, \nonumber \\ 
\mathfrak{m}_1 &=& -\frac{\eps^2}{r} \frac{\partial}{\partial r}\left( r \frac{\partial v_1 }{\partial r}  \right) + \left( \frac{m \eps}{r} \right)^2 v_1+ f''(v_0) v_1,   
\label{lin_stab_pde_deg_mobility}
\end{eqnarray} 
where $v_0(r)$ is the radially symmetric stationary state.  We solve this
system numerically, using the Chebyshev spectral collocation method
(see Appendix) with $\Delta t = 10^{-3}$ and 400 mesh points until
$t = 1/\eps^2$. The decay rate of the eigenfunction is tracked by
monitoring its maximum. The diffuse interface decay rates are scaled
with $1/\eps^2$ to compare with the sharp interface model.  The base
state that is needed for this calculation is determined \emph{a priori}
with the interface, \emph{i.e.}\ the zero contour, positioned at $r_0=0.5$.
The initial condition for the perturbation,
\begin{equation}
v_1(0,r) = \exp{\left[ {1}/({a^2 - (r_0 - r)^2 }) \right]}, 
\end{equation}  
acts approximately as a shift to the leading order shape of the inner
layer. The constant $a$ is chosen so that the support of $v_1(0,r)$
lies in the range $r>r^*$.

The results are compared in Table \ref{diffuse_sharp_deg1}.
They show that the decay rate of the azimuthal perturbation to the
radially symmetric base state obtained for $m=2$ tends to the eigenvalue
for the linearised sharp interface model \emph{with} the contribution from
nonlinear bulk diffusion, rather than to the one for pure surface
diffusion.  This confirms that \eqref{one_side_porous_medium} describes
the leading oder sharp interface evolution for the Cahn--Hilliard model
\eqref{main} correctly, and that the sharp interface motion is distinct
from the one induced by pure surface diffusion.

%%%%%%%%%%%%%%%%%%%%%%%%%%%%%%%%%%%%%%%%%%%%%%%%%%%%%%%%%%%%%%%%%%%%%%%%%%%%%%%%%%%%%%%

\section{Modifications}
\subsection{Solutions with $u>1$ for the mobility $M(u)=|1-u^2|$}
\label{init_con_sil}
As pointed out in Section~\ref{sec:axi}, solutions that have a modulus
$|u|>1$ and converge to the usual stationary Cahn--Hilliard solutions
are conceivable for the mobility $M(u)=|1-u^2|$ and are seen to arise
in numerical solutions with this mobility for appropriate initial conditions.
For this case, we can carry out the asymptotic derivations to obtain
the sharp interface limit and match the inner problem to outer solutions
on both sides of the interface, accepting thereby that the outer solution
for $u$ in $\Omega_+$ is larger than one. Otherwise the detailed derivations follow
the same pattern as in section~\ref{subsec:ord_sol} and can be found
in~\cite{mscalphalee2013}.

\begin{table}%[!h]
\centering
\begin{tabular}{| l | l | l |  l | l | l |}
\hline 
 $\eps$ & 0.01 &  0.005  & 0.002 & 0.001  & \bf Eq (\ref{solid_diffusion_two_side_decay_rate}) \\ \hline
 $\lambda_{m=2}$ & $-$144.7  & $-$146.3   & $-$147.5 & $-$147.8 & \bf $-$148.1 \\ 
 \hline
\end{tabular}
\caption{The decay rates of an azimuthal perturbation obtained by the diffuse and sharp interface models show good agreement for general initial condition not bounded between $\pm1$ and mobility $M(u) = 1-u^2$. The numerical method and discretisation parameters are the same as in Table \ref{diffuse_sharp_deg1}.  The description of the numerical approach and parameters
carries over from Table \ref{diffuse_sharp_deg1}.} 
\label{diffuse_sharp_deg2}
\end{table}

The upshot is that the sharp interface model now
has contributions from nonlinear bulk diffusion
on both sides of the interface, in addition to surface diffusion, \emph{viz.}
\begin{subequations}\label{sharp_interface_porousm}
\bl
\begin{align}
\nabla \cdot (\mu_1^\pm \nabla \mu_1^\pm) &= 0, \; \text{on} \; \Omega_{\pm}, \\    
  \mu_1^{\pm} &= \frac{2}{3} \kappa, \; \text{on} \; \zeta,  \\ 
   \nabla_n \mu_1^{+} &= 0, \; \text{on} \; \partial \Omega,\\
v_n &=  \frac{2}{3} \partial_{ss}  \kappa +  \frac{1}{4} ( \mu^+_1 \nabla_n \mu^+_1 + \mu^-_1 \nabla_n \mu^-_1), \; \text{on} \; \zeta.
\label{sipvn}
\end{align}
\el
\end{subequations}
This sharp interface model predicts an exponential decay rate of
\begin{equation}
\lambda = - \frac{2}{3} \frac{m^2(m^2-1)}{r_0^4} - \frac{1}{9} \frac{m(m^2-1)}{r_0^4}  ( \tanh (m \log r_0^{-1}) +1) \label{solid_diffusion_two_side_decay_rate}
\end{equation}
for the evolution of the perturbation to the radially symmetric stationary
state with wave number $m$. Table \ref{diffuse_sharp_deg2} shows that
equation (\ref{solid_diffusion_two_side_decay_rate}) is indeed consistent
with numerical results for the diffuse model.
As a cautionary remark, we note that we are dealing here
with a sign-changing solution of a degenerate fourth order
problem, in the sense that $1-u$ changes sign 
and the mobility degenerates. The theory
for this type of problems is still being developed
\cite{Galak10,EvansGK07,AlvarG13,BowenW06,Berni96,Galak13}.

\subsection{Degenerate biquadratic mobility} 
\label{higher_order_deg}
For the mobilities investigated so far, nonlinear bulk diffusion
enters at the same order as surface diffusion.
If we employ  $\tilde M(u) = ((1-u^2)_+)^2$, then   
\begin{equation}
j_2 = u_1 \tilde M'(u_0) \nabla_n \mu_1 = 0. 
\end{equation}
The contribution of the bulk diffusion flux to the normal velocity
of the interface is subdominant to surface diffusion and therefore 
\begin{eqnarray}\label{vnsd}
v_n  &=&
\frac{1}{3} \int_{-\infty}^{\infty}
\text{sech}^4 \rho \; \mathrm{d} \rho \; \partial_{ss} \kappa =
\frac{4}{9} \partial_{ss} \kappa.
\end{eqnarray} 
Table \ref{diffuse_sharp_deg3} shows that the decay rate 
obtained from the numerical solution of the diffuse interface model for
the degenerate biquadratic mobility
is indeed consistent with 
the predictions obtained for the sharp interface model~\eqref{vnsd}
with pure surface diffusion.

\begin{table}
\centering
\begin{tabular}{| l | l | l | l | l |}
\hline 
 $\eps$ & 0.01 &  0.005  &  0.001   & \bf Eq (\ref{pure_solid_diffusion_decay_rate})  \\ \hline
 $\lambda_{m=2}$ & $-$84.6 & $-$84.7  & $-$85.2 & $\bf $-$85.\dot{3}$ \\ 
\hline
\end{tabular}
\caption{The decay rates obtained by the diffuse interface model
for the mobility $M(u) = ((1-u^2)_+)^2$ and $|u|<1$ show good agreement with the
surface diffusion model in \eqref{pure_solid_diffusion_decay_rate}, with
$\mathfrak{M}=4/9$, as $\eps \rightarrow 0$. 
The description of the numerical approach and parameters
carries over from table~\eqref{diffuse_sharp_deg1}.}
\label{diffuse_sharp_deg3}
\end{table}

%%%%%%%%%%%%%%%%%%%%%%%%%%%%%%%%%%%%%%%%%%%%%%%%%%%%%%%%%%%%%%%%%%%%%%%%%%%%%%%%%%

\section{Conclusions}

In this paper, we have derived the sharp interface limit for a
Cahn--Hilliard model in two space dimensions with a nonlinear mobility
$M(u)=(1-u^2)_+$, and a double-well potential with minima at
$\pm 1$ for the homogeneous part of the free energy. We found that
in addition to surface diffusion, there is also a contri bution from
bulk diffusion to the interface motion which enters at the same order.
This contribution enters only from one side of the interface, whereas
for the mobility $M(u)=|1-u^2|$, solutions have also been considered for which
bulk diffusion in the sharp interface limit enters from both sides
at the same order as surface diffusion.

The situation studied here was focused on the case of
convex $\Omega_+=\{\mathbf{x}\in\Omega;\, u>0\}$
with an $O(1)$ curvature for the interace $u=0$, 
though the asymptotic analysis also remains
valid if $\Omega_+$ is the union of well-separated convex
domains.  The dynamics for concentric circles of different phases has
also been looked into \cite{mscalphalee2013}.  For the case where the
interface has turning points, the derivation needs to be revisited,
since the the location of the free boundary $\Gamma$, given by
$\rho=\sigma$ in inner coordinates about the interface, depends
on the curvature so that $|\sigma|\to\infty$ if $\kappa$ tends
to zero. Moreover, as the curvature changes sign, $\Gamma$ changes
the side of the interface.  On a different plane, it would also be
interesting to investigate the coarsening behaviour \cite{BrayE95}
for the sharp interface model \eqref{one_side_porous_medium}.
For ensembles of two or more disconnected spheres, pure surface
diffusion does not give rise to coarsening, but coarsening is expected for the
mixed surface/bulk diffusion flux in \eqref{one_side_porous_medium}.

While the Cahn--Hilliard equation \eqref{main} plays a role in some
biological models, see for example~\cite{KlappD06}, and may have
significance in modelling spinodal decomposition in porous media,
possibly with different combinations of mobilities, e.g.  $M(u) = |1-u^2|
+ \alpha (1-u^2)^2$, see \cite{mscalphalee2013}, the main motiviation for
our investigation stems from the role degenerate Cahn-Hilliard models
play as a basis for numerical simulations for surface diffusion with
interface motion driven by \eqref{vnsl}.  The upshot for the specific
combination of mobility and double well potential used in \eqref{main} is
not useful for this purpose,  since a contribution from bulk diffusion
enters at the same order.  For mobilities with higher degeneracy,
such as $M(u)=((1-u^2)_+)^2$,
this undesired effect is of higher order
and can be made arbitrarily small, at least
in principle, by reducing $\eps$. 
Nevertheless, for finite $\eps$, it is still present and a
cumulative effect may arise for example through a small but persistent
coarsening of phase-separated domains. 

A range of alternatives can be found in the
literature, in particular using the combination of $M=(1-u^2)_+$ or $M=|1-u^2|$ with the
logarithmic or with the double obstacle potential \cite{cahn1996cahn}. 
These combinations force the order
parameter $u$ to be equal to or much closer to $\pm 1$ away from the interface, 
thus shutting out the bulk diffusion more effectively.  
Numerical methods have been developed for these
combinations and investigated in
the literature, see for example~\cite{BarreB02, BarreBG02, BarreBG98,
BarreBG99, BarreGN07, GarckNS07, BanasNN13}.  Other approaches that have
been suggested include a dependence of the mobility on the gradients
of the order parameter \cite{mahadevan1999phase}, tensorial mobilities
\cite{gugenberger2008comparison}, or singular expressions for the chemical
potential \cite{RaRV06}.

As a final remark, we note that many analytical questions remain open.
For example, the existence of solutions that preserve the property that $|u|>1$
in some parts of $\Omega$ has not been shown.
Also, the approximation or \eqref{main} by a free boundary
problem \eqref{chefb} should be investigated systematically using
$b=\text{min } (1-|u|)>0$ as a small parameter, in the spirit of what
was done, for example, in \cite{KingB01} for the precursor model of a
spreading droplet. The conditions at the free boundary $\Gamma$ could
then be recovered from matching to an inner solution.  If $b\to 0$
in finite time, 
the effect of the ``precursor'' regularisation is lost and either
the regularising effect implicit in the numerical discretisation or
any explicit regularisation that is used (e.g., the one
suggested in \cite{EllioG96}) have to be taken into account. It
would be interesting to see for which regularisations the conditions in \eqref{chefb_freebcs}
are recovered. We note, however, that the evolution of the leading order
sharp interface model in $\Omega_-$ is usually insensitive to the conditions
imposed at $\Gamma$. 

\section{Appendix: Numerical Methods} 
We numerically  solved the radially symmetric counterpart to \eqref{main} in polar coordinates
without an explicit regularsisation (such as the on used in \cite{EllioG96}) 
via a Chebyshev spectral
collocation method in space and semi-implicit
time-stepping, using a linearised convex splitting scheme to treat $f$. 
For details on spectral methods in general, we refer the reader
to the references \cite{trefethen2000spectral,trefethen2013approximation}.
We also split the mobility as $M(u) \equiv (M(u) - \theta) + \theta$, to evaluate $(M(u) - \theta)$ at the previous time step whilst solving the remaining $\theta$ portion at the next time step,
which improved the stability.  We choose $\theta = 0.01 \eps$ in our simulations.
Varying $\theta$ confirmed that the results did not sensitively depend on its
value provided it was $O(\eps)$.

As the Chebyshev--Lobatto points are scarcest in the middle of the
domain, we resolve the interior layer by introducing a non-linear map
$x \in [-1,1] \mapsto r \in
[0,1]$, as suggested in \cite{boyd1992arctan},
$
r =({1}/{2}) +\mathrm{arctan} \left(\delta \tan \pi x/2 \right)/\pi,
$
where $0<\delta<1$ is a parameter that determines the degree of stretching of the interior domain, with a smaller value of $\delta$ corresponding to greater degree of localisation of mesh points about the centre of the domain. In this paper, we general set  $\delta = 10 \eps$.
This choice of $\delta$ is guided by numerical experiments, which show that further increase in the number of mesh points does not alter the stationary solution. 
Moreover, since $r=0$ is a regular singular point, 
we additionally map the domain $r\in[0,1]$ linearly onto a truncated domain $[10^{-10},1]$. Again, we verified that varying the truncation parameter did not
affect the numerical results.
Unless otherwise stated, the numerical simulations reported in the paper
are done with 400 collocation points and timestep $\Delta t = 10^{-3}$.

The linearised phase-field models were solved using the same method, with a
base state that was obtained from a preceding run and then ``frozen'' in time, \emph{i.e.}\
not co-evolved with the perturbation.

%%%%%%%%%%%%%%%%%%%%%%%%%%%%%%%%%%%%%%%%%%%%%%%%%%%%%%%%%%%%%%%%%%%%%%%
\bibliographystyle{abbrv}
\bibliography{pf}

\begin{thebibliography}{10}

\bibitem{AbelsGG12}
H.~Abels, H.~Garcke, and G.~Gr\"un.
\newblock Thermodynamically consistent, frame indifferent diffuse interface
  models for incompressible two-phase flows with different densities.
\newblock {\em Mathematical Models and Methods in Applied Sciences},
  22(03):1150013, 2012.

\bibitem{AbelsR09}
H.~Abels and M.~R{\"o}ger.
\newblock Existence of weak solutions for a non-classical sharp interface model
  for a two-phase flow of viscous, incompressible fluids.
\newblock {\em Annales de {l'Institut} Henri Poincare (C) Non Linear Analysis},
  26(6):2403--2424, Nov. 2009.

\bibitem{alikakos1994convergence}
N.~D. Alikakos, P.~W. Bates, and X.~Chen.
\newblock Convergence of the {{Cahn-Hilliard}} equation to the {Hele-Shaw}
  model.
\newblock {\em Archive for Rational Mechanics and Analysis}, 128(2):165--205,
  1994.

\bibitem{AlvarG13}
P.~Alvarez-Caudevilla and V.~A. Galaktionov.
\newblock Well-posedness of the cauchy problem for a fourth-order thin film
  equation via regularization approaches.
\newblock {\em {arXiv:1311.0712} [math]}, Nov. 2013.

\bibitem{BanasNN13}
L.~Bana\u{s}, A.~Novick-Cohen, and R.~N\"{u}ernberg.
\newblock The degenerate and non-degenerate deep quench obstacle problem: A
  numerical comparison.
\newblock {\em Networks and Heterogeneous Media}, 8(1):37--64, Mar. 2013.

\bibitem{BarreB02}
J.~W. Barrett and J.~F. Blowey.
\newblock Finite element approximation of a degenerate
  {Allen--Cahn/Cahn--Hilliard} system.
\newblock {\em {SIAM} Journal on Numerical Analysis}, 39(5):1598--1624, Jan.
  2002.

\bibitem{BarreBG98}
J.~W. Barrett, J.~F. Blowey, and H.~Garcke.
\newblock Finite element approximation of a fourth order nonlinear degenerate
  parabolic equation.
\newblock {\em Numerische Mathematik}, 80(4):525--556, Oct. 1998.

\bibitem{BarreBG99}
J.~W. Barrett, J.~F. Blowey, and H.~Garcke.
\newblock Finite element approximation of the {Cahn--Hilliard} equation with
  degenerate mobility.
\newblock {\em {SIAM} Journal on Numerical Analysis}, 37(1):286--318, Jan.
  1999.

\bibitem{BarreBG02}
J.~W. Barrett, J.~F. Blowey, and H.~Garcke.
\newblock On fully practical finite element approximations of degenerate
  {Cahn-Hilliard} systems.
\newblock {\em {ESAIM:} Mathematical Modelling and Numerical Analysis},
  35(4):713--748, Apr. 2002.

\bibitem{BarreGN07}
J.~W. Barrett, H.~Garcke, and R.~N{\"u}rnberg.
\newblock A phase field model for the electromigration of intergranular voids.
\newblock {\em Interfaces and Free Boundaries}, 9:171{\textendash}210, 2007.

\bibitem{Berni96}
F.~Bernis.
\newblock Finite speed of propagation and continuity of the interface for thin
  viscous flows.
\newblock {\em Advances in Differential Equations}, 1(3):337--368, 1996.

\bibitem{bhate2000diffuse}
D.~N. Bhate, A.~Kumar, and A.~F. Bower.
\newblock Diffuse interface model for electromigration and stress voiding.
\newblock {\em Journal of Applied Physics}, 87(4):1712--1721, 2000.

\bibitem{BowenW06}
M.~Bowen and T.~P. Witelski.
\newblock The linear limit of the dipole problem for the thin film equation.
\newblock {\em {SIAM} Journal on Applied Mathematics}, 66(5):1727--1748, May
  2006.

\bibitem{boyd1992arctan}
J.~P. Boyd.
\newblock The arctan/tan and {Kepler-Burgers} mappings for periodic solutions
  with a shock, front, or internal boundary layer.
\newblock {\em Journal of Computational Physics}, 98(2):181--193, 1992.

\bibitem{BrayE95}
A.~J. Bray and C.~L. Emmott.
\newblock {Lifshitz-Slyozov} scaling for late-stage coarsening with an
  order-parameter-dependent mobility.
\newblock {\em Physical Review B}, 52(2):R685--R688, July 1995.

\bibitem{CahnH71}
J.~Cahn and J.~Hilliard.
\newblock Spinodal decomposition: A reprise.
\newblock {\em Acta Metallurgica}, 19(2):151--161, Feb. 1971.

\bibitem{CahnT94}
J.~Cahn and J.~Taylor.
\newblock Surface motion by surface diffusion.
\newblock {\em Acta Metallurgica et Materialia}, 42(4):1045--1063, Apr. 1994.

\bibitem{cahn1996cahn}
J.~W. Cahn, C.~M. Elliott, and A.~Novick-Cohen.
\newblock The {Cahn-Hilliard} equation with a concentration dependent mobility:
  motion by minus the laplacian of the mean curvature.
\newblock {\em European Journal of Applied Mathematics}, 7(3):287--302, 1996.

\bibitem{CahnH58}
J.~W. Cahn and J.~E. Hilliard.
\newblock Free energy of a nonuniform system. i. interfacial free energy.
\newblock {\em The Journal of Chemical Physics}, 28(2):258, 1958.

\bibitem{CenicG13}
H.~D. Ceniceros and C.~J. Garc{\'\i}a-Cervera.
\newblock A new approach for the numerical solution of diffusion equations with
  variable and degenerate mobility.
\newblock {\em Journal of Computational Physics}, 2013.

\bibitem{chen2002phase}
L.~Chen.
\newblock Phase-field models for microstructure evolution.
\newblock {\em Annual review of materials research}, 32(1):113--140, 2002.

\bibitem{DaiD12}
S.~Dai and Q.~Du.
\newblock Motion of interfaces governed by the {Cahn--Hilliard Equation} with
  highly disparate diffusion mobility.
\newblock {\em {SIAM} Journal on Applied Mathematics}, 72(6):1818--1841, Nov.
  2012.

\bibitem{EllioG96}
C.~M. Elliott and H.~Garcke.
\newblock On the {Cahn-Hilliard} equation with degenerate mobility.
\newblock {\em SIAM Journal on Mathematical Analysis}, 27(2):404--423, 1996.

\bibitem{EvansGK07}
J.~D. Evans, V.~A. Galaktionov, and J.~R. King.
\newblock Source-type solutions of the fourth-order unstable thin film
  equation.
\newblock {\em European Journal of Applied Mathematics}, 18(3):273--321, 2007.

\bibitem{Galak10}
V.~A. Galaktionov.
\newblock Very singular solutions for thin film equations with absorption.
\newblock {\em Studies in Applied Mathematics}, 124(1):39{\textendash}63, 2010.

\bibitem{Galak13}
V.~A. Galaktionov.
\newblock On {Oscillations} of {Solutions} of the {Fourth}-{Order} {Thin}
  {Film} {Equation} {Near} {Heteroclinic} {Bifurcation} {Point}.
\newblock {\em arXiv:1312.2762 [math]}, 2013.

\bibitem{GarckNS07}
H.~Garcke, R.~N{\"u}rnberg, and V.~Styles.
\newblock Stress- and diffusion-induced interface motion: Modelling and
  numerical simulations.
\newblock {\em European Journal of Applied Mathematics}, 18(6):631--657, 2007.

\bibitem{giacomin1996exact}
G.~Giacomin and J.~L. Lebowitz.
\newblock Exact macroscopic description of phase segregation in model alloys
  with long range interactions.
\newblock {\em Physical Review Letters}, 76(7):1094, 1996.

\bibitem{giacomin1997phase}
G.~Giacomin and J.~L. Lebowitz.
\newblock Phase segregation dynamics in particle systems with long range
  interactions {I}: Macroscopic limits.
\newblock {\em Journal of Statistical Physics}, 87(1-2):37--61, 1997.

\bibitem{giacomin1998phase}
G.~Giacomin and J.~L. Lebowitz.
\newblock Phase segregation dynamics in particle systems with long range
  interactions {II}: Interface motion.
\newblock {\em SIAM Journal on Applied Mathematics}, 58(6):1707--1729, 1998.

\bibitem{giacomin2000macroscopic}
G.~Giacomin, J.~L. Lebowitz, and R.~Marra.
\newblock Macroscopic evolution of particle systems with short-and long-range
  interactions.
\newblock {\em Nonlinearity}, 13(6):2143, 2000.

\bibitem{Glasn03}
K.~Glasner.
\newblock A diffuse interface approach to {Hele{\textendash}Shaw} flow.
\newblock {\em Nonlinearity}, 16(1):49, 2003.

\bibitem{gugenberger2008comparison}
C.~Gugenberger, R.~Spatschek, and K.~Kassner.
\newblock Comparison of phase-field models for surface diffusion.
\newblock {\em Physical Review E}, 78(1):016703, 2008.

\bibitem{jiang2012phase}
W.~Jiang, W.~Bao, C.~V. Thompson, and D.~J. Srolovitz.
\newblock Phase field approach for simulating solid-state dewetting problems.
\newblock {\em Acta Materialia}, 60(15):5578--5592, 2012.

\bibitem{KingB01}
J.~R. King and M.~Bowen.
\newblock Moving boundary problems and non-uniqueness for the thin film
  equation.
\newblock {\em European Journal of Applied Mathematics}, 12(03):321--356, 2001.

\bibitem{kltahara1978kinetic}
K.~Kitahara and M.~Imada.
\newblock On the kinetic equations for binary mixtures.
\newblock {\em Progress in Theoretical Physics Suppliment}, 64:65--73, 1978.

\bibitem{KlappD06}
I.~Klapper and J.~Dockery.
\newblock Role of cohesion in the material description of biofilms.
\newblock {\em Physical Review E}, 74(3), Sept. 2006.

\bibitem{korzec2008stationary}
M.~D. Korzec, P.~L. Evans, A.~M{\"u}nch, and B.~Wagner.
\newblock Stationary solutions of driven fourth-and sixth-order
  {{Cahn-Hilliard}-type} equations.
\newblock {\em SIAM Journal on Applied Mathematics}, 69(2):348--374, 2008.

\bibitem{langer1983}
C.~G. Lange.
\newblock On spurious solutions of singular perturbation problems.
\newblock {\em Studies in Applied Mathematics}, 68:227--257, 1983.

\bibitem{mscalphalee2013}
A.~A. Lee.
\newblock {On the Sharp Interface Limits of the {Cahn-Hilliard} Equation}.
\newblock {M.Sc. thesis}, University of Oxford, 2013.

\bibitem{LuGBK07}
H.-W. Lu, K.~Glasner, A.~L. Bertozzi, and C.-J. Kim.
\newblock A diffuse-interface model for electrowetting drops in a hele-shaw
  cell.
\newblock {\em Journal of Fluid Mechanics}, 590:411--435, 2007.

\bibitem{mahadevan1999phase}
M.~Mahadevan and R.~M. Bradley.
\newblock Phase field model of surface electromigration in single crystal metal
  thin films.
\newblock {\em Physica D: Nonlinear Phenomena}, 126(3):201--213, 1999.

\bibitem{Mulli57}
W.~W. Mullins.
\newblock Theory of thermal grooving.
\newblock {\em Journal of Applied Physics}, 28:333, 1957.

\bibitem{mullins1963morphological}
W.~W. Mullins and R.~F. Sekerka.
\newblock Morphological stability of a particle growing by diffusion or heat
  flow.
\newblock {\em Journal of Applied Physics}, 34(2):323--329, 1963.

\bibitem{Nieth95}
B.~S. Niethammer.
\newblock Existence and uniqueness of radially symmetric stationary points
  within the gradient theory of phase transitions.
\newblock {\em European Journal of Applied Mathematics}, 6(01), Feb. 1995.

\bibitem{pego1989front}
R.~L. Pego.
\newblock Front migration in the nonlinear {Cahn-Hilliard} equation.
\newblock {\em Proceedings of the Royal Society of London. A. Mathematical and
  Physical Sciences}, 422(1863):261--278, 1989.

\bibitem{provatas2010phase}
N.~Provatas and K.~Elder.
\newblock {\em Phase-field methods in materials science and engineering}.
\newblock Wiley Interscience, 2010.

\bibitem{PuriBL97}
S.~Puri, A.~J. Bray, and J.~L. Lebowitz.
\newblock Phase-separation kinetics in a model with order-parameter-dependent
  mobility.
\newblock {\em Physical Review E}, 56(1):758--765, July 1997.

\bibitem{ratz2006surface}
A.~R{\"a}tz, A.~Ribalta, and A.~Voigt.
\newblock Surface evolution of elastically stressed films under deposition by a
  diffuse interface model.
\newblock {\em Journal of Computational Physics}, 214(1):187--208, 2006.

\bibitem{RaRV06}
A.~R{\"a}tz, A.~Ribalta, and A.~Voigt.
\newblock Surface evolution of elastically stressed films under deposition by a
  diffuse interface model.
\newblock {\em Journal of Computational Physics}, 214(1):187--208, May 2006.

\bibitem{SibleNK13}
D.~N. Sibley, A.~Nold, and S.~Kalliadasis.
\newblock Unifying binary fluid diffuse-interface models in the sharp-interface
  limit.
\newblock {\em Journal of Fluid Mechanics}, 736:5--43, 2013.

\bibitem{TayloC94}
J.~E. Taylor and J.~W. Cahn.
\newblock Linking anisotropic sharp and diffuse surface motion laws via
  gradient flows.
\newblock {\em Journal of Statistical Physics}, 77(1-2):183--197, Oct. 1994.

\bibitem{TorabL12}
S.~Torabi and J.~Lowengrub.
\newblock Simulating interfacial anisotropy in thin-film growth using an
  extended cahn-hilliard model.
\newblock {\em Physical Review E}, 85(4):041603, Apr. 2012.

\bibitem{TorabLVW09}
S.~Torabi, J.~Lowengrub, A.~Voigt, and S.~Wise.
\newblock A new phase-field model for strongly anisotropic systems.
\newblock {\em Proceedings of the Royal Society A: Mathematical, Physical and
  Engineering Science}, 465(2105):1337--1359, May 2009.

\bibitem{trefethen2000spectral}
L.~N. Trefethen.
\newblock {\em Spectral methods in MATLAB}, volume~10.
\newblock {SIAM}, 2000.

\bibitem{trefethen2013approximation}
L.~N. Trefethen.
\newblock {\em Approximation theory and approximation practice}.
\newblock {SIAM}, 2013.

\bibitem{GemmeBP05}
S.~van Gemmert, G.~T. Barkema, and S.~Puri.
\newblock Phase separation driven by surface diffusion: A monte carlo study.
\newblock {\em Physical Review E}, 72(4):046131, Oct. 2005.

\bibitem{WolteBP06}
J.~K. Wolterink, G.~T. Barkema, and S.~Puri.
\newblock Spinodal decomposition via surface diffusion in polymer mixtures.
\newblock {\em Physical Review E}, 74(1):011804, July 2006.

\end{thebibliography}

%%%%%%%%%%%%%%%%%%%%%%%%%%%%%%%%%%%%%%%%%%%%%%%%%%%%%%%%%%%%%%%%%%%%%%%%%

\end{document}